\documentclass[twocolumn,secnumarabic,amssymb, nobibnotes, aps, prb]{revtex4-1}

\usepackage{amsmath,amssymb}
\usepackage{graphicx}
\usepackage{footnote}
\usepackage{subfigure}

\begin{document}

\title{Detecting a preformed pair phase: The response to a pairing forcing field}%

\author{A.~Tagliavini,$^{1}$ M.~Capone,$^{2}$ and A.~Toschi,$^{1}$}%
%\email[REVTeX Support: ]{revtex@aps.org}
\affiliation{$^1$ Institute for Solid State Physics, Vienna University of Technology, 1040 Vienna, Austria }
\affiliation{$^2$ International School for Advanced Studies (SISSA), Via Bonomea 265, I-34136, Trieste, Italy}
\date{\small\today}%

\begin{abstract}
{The normal state of strongly coupled superconductors is characterized by the presence of ``preformed'' Cooper pairs well above the superconducting critical temperature. In this regime, the electrons are paired, but they lack the phase coherence necessary for superconductivity. The existence of preformed pairs implies the existence of a characteristic energy scale associated to a pseudogap. 
Preformed pairs are often invoked to interpret systems where some signatures of pairing are present without actual superconductivity, but  an unambiguous theoretical characterization of a preformed-pair system is still lacking. To fill this gap, we consider the response to an external pairing field of an attractive Hubbard model, which hosts one of the cleanest realizations of a preformed pair phase, and a repulsive model where $s-$wave superconductivity can not be realized. Using dynamical mean-field theory to study this response, we identify the characteristic features which distinguish the reaction of a preformed pair state from a normal metal without any precursor of pairing. The theoretical detection of preformed pairs is associated with the behavior of the {\sl second} derivative of the order parameter with respect to the external field, as confirmed by analytic calculations in limiting cases.
Our findings provide a solid testbed for the interpretation of state-of-the-art calculations for the normal state of the doped Hubbard model in terms of  $d-$wave preformed pairs and, in perspective, of non-equilibrium experiments in high-temperature superconductors.}
\pacs{}
\end{abstract}

\maketitle

\section{Introduction}
In many complex materials and quantum systems we witness the persistence of fingerprints of superconductivity well above the critical temperature and clearly distinct from fluctuation phenomena. This often leads to a possible interpretation in terms of  electron pairs which are formed at very large temperature but they can condense only at a much lower critical temperature due to the phase fluctuations of their wavefunction. Yet, the unambiguous detection of preformed pairs is elusive, as it does not correspond to an actual phase transition and it can not be unambiguously associated with a direct observable quantity. 

The prototypical realization of this physics takes place in model systems with strong pairing interaction, which drives the formation of tightly bound pairs with a reduced phase coherence. In this regime, superconductivity occurs as a Bose-Einstein condensation (BEC) of composite bosons formed by the bound pairs of fermions. When the pairing strength is tuned from weak to strong coupling one observes a continuous crossover from the familiar BCS \cite{Bardeen1957}  pairing to this regime.

%The formation of bound pairs is a central aspect of the physics of fermions in the presence of a strong attractive interaction. In fact, in this limiting case, the physical description becomes particularly transparent:
%Fermions are bound in bosonic pairs because of the dominant attractive interaction, and, if parameters (such dimensions, temperature, etc.) allow, they might undergo a Bose-Einstein condensation (BEC) and display superfluidity or superconductivity.  The latter deviates from the ``conventional'' BCS\cite{Bardeen1957} one, typical of the weak-coupling regime, because its onset is entirely controlled by the phase coherence of the wavefunctions of the electronic pairs. Hence, in the BEC regime, the critical temperature ($T_c$) scales with the superfluid stiffness ($D_s$) and the superconducting phase is stabilized by a gain of kinetic, rather than potential, energy. 

This BCS-BEC crossover\cite{BCSBE,BCSBE2,BCSBE3,BCSBE4,BCSBE5} has been intensively studied, both in the context of cold atoms trapped in optical lattices\cite{Greiner2009} and in high temperature superconductivity, where a preformed pair regime has been invoked\cite{preform1,preform2,preform3,preform4, preform5} for the pseudogap state\cite{Timusk1999,Campuzano2004} of underdoped cuprates.

%and recently also in some of their nickelate analogue\cite{Uchida2011}.

In this work we use the attractive Hubbard model as a theoretical device to set a {\it{practical protocol}} to confirm or disprove the existence of preformed pairs in a specific system under analysis. Comparing regimes where $s$-wave preformed pairs are certainly present or certainly absent, we identify which properties of the system are so sensitive to their presence to be exploited for their detection. Such an identification will also be applicable to interpret existing analyses of the pseudogap-phase in the cuprates\cite{Gull_Millis2012,Merino2014,Gunnarsson2015}.

%A clear-cut identification of the hallmarks of the response of a preformed-pair phase to an external pairing field, however, can be only carried out via a comparative study against cases where the preformed pairs are surely absent. In our DMFT study, this is easily realized by considering the effect of the $s-$wave pairing field on a repulsive ($U >0$) Hubbard model: The single-site DMFT  only allows for $s-$wave pairing fluctuations, which are suppressed by the local repulsion.   
 
We have structured our paper as follows: In Sec. II, we briefly discuss the modellization of the problem, in terms of the single band (attractive) Hubbard Hamiltonian, and briefly review some of the previous DMFT studies in absence of an external field. In Sec. III, we report our DMFT results in presence of a forcing field at different temperatures and interactions, comparing explicitly the attractive and repulsive models. 
The physical interpretation of our numerical results in terms of the underlying ground state properties is given in Sec. IV. In Sec. V we discuss the implication of this criterion in a broader context, while the final Sec. VI presents our conclusions.
\section{Modelization of the problem}
\label{Model}

Throughout this paper we will consider a simple Hubbard model in presence of an external field driving an s-wave superconducting order parameter
\begin{multline}
\label{complete_H_eq}
H = -t \ \sum_{<ij> \sigma} c^{\dagger}_{i \sigma} c_{j \sigma} + U \sum_{i} n_{i\uparrow} n_{i\downarrow} \\
 - \mu \sum_{i \sigma} n_{i \sigma}  - \eta \sum_{i} (c^{\dagger}_{i\uparrow} c^{\dagger}_{i\downarrow} + h.c),
\end{multline}
Here, $t$ represents the nearest-neighbor amplitude, $\mu$ is the chemical potential, and the effective interaction $U$ is negative for the {\sl attractive} and positive for the {\sl repulsive} Hubbard model.
%\footnote{We recall that the repulsive and the attractive cases can be mapped one into the other by performing a unitary transformation of the fermionic operators [\onlinecite{BCSBE}]. In this way the physics of one system can be easily `translated' into the physics of its correspondent counterpart: for example, at the level of the ground state and at half-filling, the unperturbed ($\eta=0$) systems exhibit a antiferromagnetic (AFM) ground state for $U>0$ which becomes an s-wave superconducting ground state degenerate with a charge density wave (CDW) for $U<0$.}. 
The last contribution in Eq. (\ref{complete_H_eq}) represents the coupling of the system to a forcing, time-independent pairing field $\eta$, which is assumed to be positive and isotropic ({\sl $s$-wave}). 

In more than two dimensions the attractive Hubbard model displays a low-temperature $s$-wave superconductivity, smoothly evolving from a weak-coupling (BCS) regime to a strong-coupling BEC regime with increasing $U$\cite{Toschi2005a}. In the latter regime, pairs are formed at a very high temperature of order U, while they can only condense at a much lower temperature $T = T_c \propto \frac{1}{U}$ because of the large phase fluctuations which contrast the formation of a coherent condensate\cite{BCSBE}. In the BCS regime, superconductivity is stabilized by a potential energy gain, while the superconductor has a (slightly) higher kinetic energy than the normal state. In the BEC regime the energetic balance is the opposite: The superconductor is stabilized by a kinetic energy gain with a slight potential energy loss w.r.t the normal state\cite{Toschi2005a}.

Given the $s$-wave nature of the pairs and the local nature of the interactions, much of the physics for $d > 2$ can be well captured by dynamical mean field theory (DMFT)\cite{Georges1996}, an approach where spatial fluctuations are frozen, but the local dynamics is included non-perturbatively at every value of the interaction strength. DMFT becomes formally exact in the limit of infinite coordination\cite{Metzner1989} of the lattice but it can be used as an approximation in finite dimensions, where it provides a fully non-perturbative description. This represents a major advantage to analyze weak and strong coupling regimes on equal footing. Previous DMFT studies\cite{DMFTUneg,Toschi2005a,Toschi2005b,DMFTUneg2} have focused on spectral, thermodynamic properties, and even to some non-equilibrium properties\cite{Amaricci2016}. Here we consider a different aspect, namely the response to an external stimulus which drives a superconducting s-wave pairing, also beyond the linear-response regime.
% For our purpose, however, this is not sufficient: We aim at a theoretical investigation able of selectively detecting the presence or the absence of preformed pairs, and their promptness to sustain a superconducting order under a proper stimulation. This goal can be achieved by studying in detail the effect of the general response (i.e., {\sl not} limited to the linear regime) to an external perturbation, which couples with the preformed pairs, if they are present:  Such perturbation can be readily identified with an $s-$wave {\sl pairing} field $\eta$, i.e., the one which induces a diverging superconducting linear response at $T=T_c$. 
% We note that a calculation in presence of a d-wave pairing forcing field has been recently carried on \cite{Gull_Millis2012} via the dynamical cluster approximation (DCA)\cite{DCArev} of the repulsive Hubbard model: We will explicitly discuss the relation of these results with our work in the last part of our paper. 

%As we focus on preformed pair states with the most basic ($s$-wave) symmetry, we perform our calculations with the single-site dynamical mean-field theory (DMFT)[\onlinecite{Georges1996}], which allows for an isotropic (spontaneous or induced) superconducting symmetry breaking. We note, however, that some general conclusions of our study will be applicable to the cases of other singlet pairings, i.e., for the detection of preformed pairs with $d-$wave symmetry within DCA, which is highly relevant for, e.g., the cuprate physics. 

\begin{figure}[t!]
\includegraphics[width=0.45\textwidth]{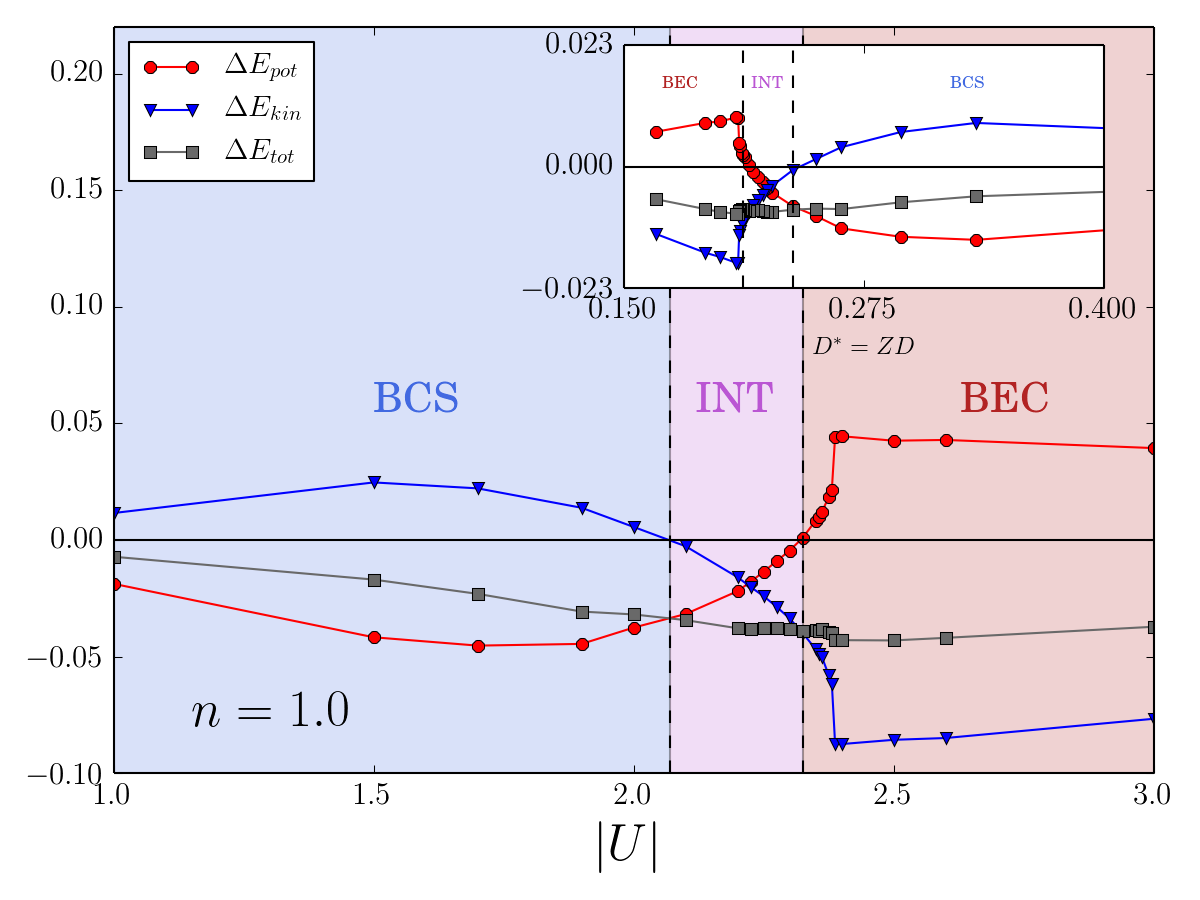}
\includegraphics[width=0.50\textwidth]{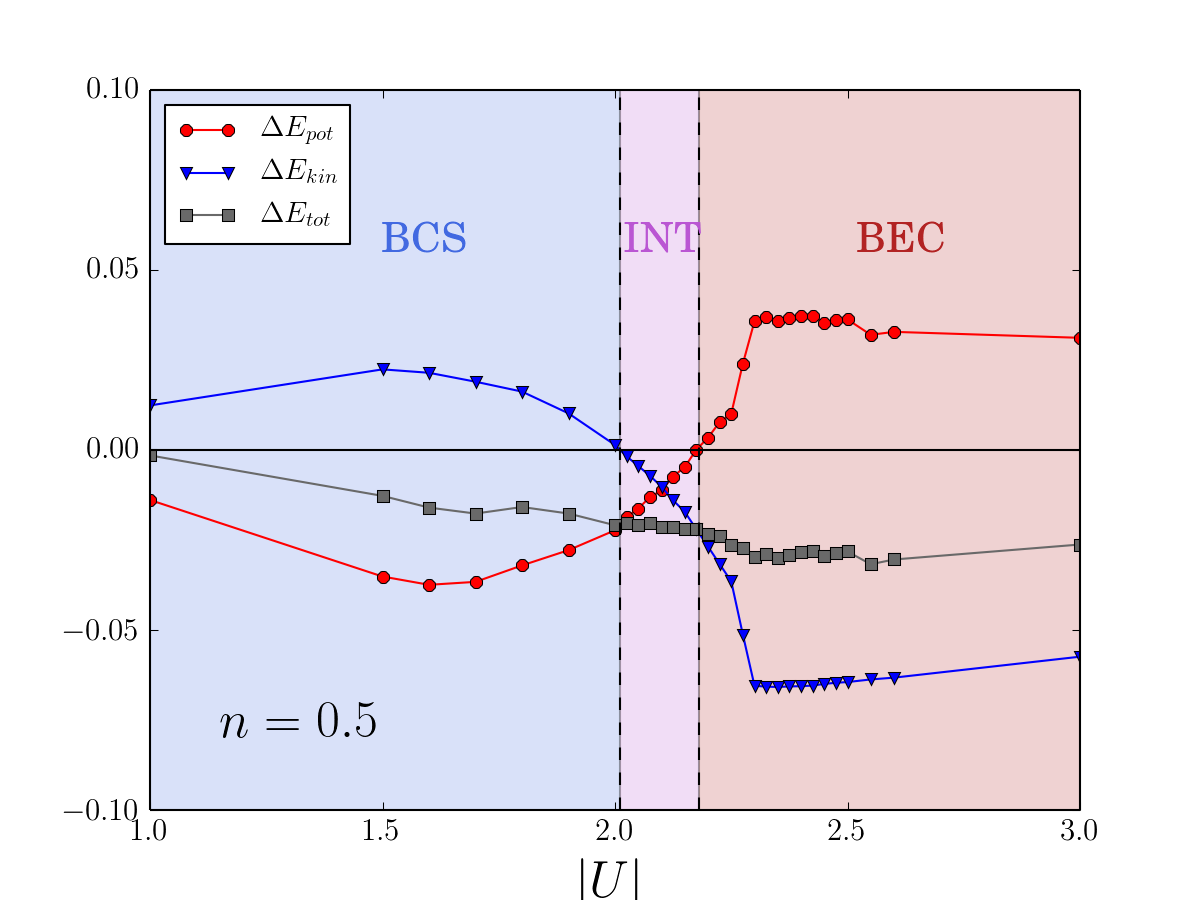}
\caption{{\small Energy balance $\Delta E_{tot} = E_S - E_N$ (and its kinetic and potential components) computed in the superconducting region at $\beta= 50 D^{-1}$ as a function of the attractive interaction $U$. The upper panel refers to the half filled ($n=1$) case, while the lower panel refers to the electron density $n=0.5$. Inset: The same but for the Hamiltonian of Eq. (\ref{resc_H_eq}), where the energies have been scaled in order to keep the attractive interaction constant ($\lambda = -0.5)$ and let the bandwidth vary as a more realistic effect of the correlation and/or doping.}}
\label{fig:energy_balance}
\end{figure} 

For the sake of definiteness we consider a semicircular density of states $N(\epsilon)= \frac{2}{\pi D^2} \sqrt{D^2 -\epsilon^2} $  ($D$ being its half-bandwidth) which is suitable to represent a  finite-bandwidth system in DMFT. To solve the auxiliary impurity problem of DMFT we adopted an exact diagonalization (ED) solver with $n_s = 5$ sites (one impurity and $n_b = 4$ bath electronic sites), and tested the stability of the results by increasing the number of sites in the most relevant intermediate coupling/low-temperature regime.  Obviously, due to the external pairing field in Eq. (\ref{complete_H_eq}),  the DMFT treatment has to be extended to the broken-symmetry phase, by recasting DMFT in Nambu formalism (see, e.g., [\onlinecite{Toschi2005b}]).

In this section we set up the stage by presenting refined results for the unperturbed attractive Hubbard model. In this way we identify concretely weak-, intermediate- and strong-coupling regions whose definition will be helpful to guide the discussion of the following sections.

%Before entering the main part of our work, i.e. the effect of the forcing pairing field $\eta$, it should be recalled that several DMFT calculations for the attractive Hubbard model {\sl for} $\eta =0$ have already been performed in the past to investigate the BCS-BEC crossover both in the disordered (normal) and in the long-range ordered ($s$-wave superconducting) phase.
%In particular, in the resulting phase diagram, a continuous evolution from the BCS to the BE superconductivity through an intermediate regime has been identified by studying the energetic mechanism that stabilizes the ordered phase: One distinguishes the cases where the superconductivity is driven by a gain in  potential (BCS), kinetic energy (BEC) or both (intermediate) [\onlinecite{Toschi2005a}]. 
%Hence,  in order to set the stage for studying the action of the pairing field, we briefly summarize what can be regarded as our staring point: The main DMFT results in absence of a pairing field ($\eta =0$). 
%We will focus here on the general phase diagram for different electron densities, whose qualitative structure largely coincides with the literature results (though our calculations also allows for  quantitative refinements of the borders between the different regimes w.r.t. the previous estimates). 

Fig.~\ref{fig:energy_balance} shows the energy difference between the superconducting and the normal state $\Delta E_{tot} = E_{S} - E_{N}$ (resolved in its kinetic and potential energy components)  in different interaction regimes for two significant choices of the electron density: $n=1$ (half-filling) and $n=0.5$ (quarter-filling). We consider a low value of the temperature 
$\beta = 50 D^{-1}$ which is significantly below $T_c$ for the broad range of $\vert U \vert$ used in the figure.
Moreover, we verified that discretization effects associated with a finite bath size of the ED solver are negligible.

%%%%%%%%%%%%%%%%%%%%%%%%%%%%
Our findings are summarized by the different colors in the diagrams, which mark the different regimes (BCS: blue; intermediate: violet; BEC: red). Fig.~1 shows that a qualitative change of the energetic balance w.r.t. BCS only takes place when $U\simeq2D$, where a narrow intermediate region, in which the superconductor gains both potential and kinetic energy starts. At $U\simeq2.5D$ a BEC regime establishes. This evolution of the energetic balance tracks the progressive formation of preformed pairs in the normal state.

%The kinetic energy gain which is observed for higher $U$ is directly associated to preformed Cooper pairs in the normal state, whose Bose-Einstein condensation helps to stabilize the superconducting phase. The preformed pair physics becomes more limpid only for $U \simeq 2.5 D$, where the Bose-Einstein condensation remains the {\sl only} process that stabilizes the ordered phase.

These results provide a more accurate determination of the boundaries found in Ref.~[\onlinecite{Toschi2005a}].  We also notice that the results for $n=1$, and $n=0.5$ are remarkably similar. This observation shows how weakly the physics of the attractive Hubbard model depends on doping. Half-filling, thus, does not represent a special point for the superconducting solution, despite of its peculiar degeneracy with the charge density wave. This justifies the use of the half filled case for the  general analysis in the next section, which is numerically less costly due to particle-hole symmetry. 
Analogous DMFT characterizations also hold for magnetic phases, in particular for the ``sibling'' crossover from a Slater to an Heisenberg antiferromagnet, as explicitly shown by recent DMFT\cite{Taranto2012} and DCA\cite{Gull2008} results.

%Finally, it is worth recalling that the importance of studying the energetics of a superconductor, as presented above, goes well beyond an ``academic'' classification of the different superconducting regimes. 
In this work the attractive Hubbard is not introduced as a microscopic description of any realistic material, but as a simple tool for the detection of preformed pairs. However, the energetic analysis we just summarized was suggested as a possible explanation of the spectral weight changes observed in the optical conductivity on the cuprates\cite{Molegraaf2002,Santander2002,Santander2003,Carbotte2006,Qazil2009,Benfatto2004, Toschi2005c,Baldassarre2008,Toschi2008,Nicoletti2010,Deutscher2005,Carbone2006}.
Evidently any attempt in this direction must include at least qualitatively the effect of strong repulsive correlations, which mainly control the doping dependence of the cuprate phase diagram. Thus, if one wanted to use Eq.~(\ref{complete_H_eq}) with $U = - \lambda < 0$ for a rough investigation of these specific features in the cuprate physics, one should account for the doping dependence of the quasiparticle properties. This can be achieved by renormalizing the kinetic term by means of the quasiparticle weight $Z$, while leaving at first approximation the effective attractive interaction unchanged:

%This aspect of the controversial physics of material class may be captured by simplified approaches (e.g. slave boson theories [\onlinecite{Wlfle1995}], or low-energy strongly correlated superconductivity scenarios [\onlinecite{Capone2002}]) in which a low-energy picture is built in terms of strongly renormalized quasiparticles. Eventually these already renormalized excitations are the ones, which collectively experience an effective attraction.

%However, a conceptual connection with the gross structure of our phase-diagrams and the physics of strongly correlated superconductors can be only established by fully neglecting the microscopic origin of the effective attraction, taken as an oversimplified, constant local term in Eq. (\ref{complete_H_eq}), and assumed {\sl unaffected} by the local Coulomb repulsion. In such framework,  an evolution from BCS to BEC-like features in the physics can be ascribed to the unbalance of the ratio between the constant attraction and an effective bandwidth, progressively narrowed  by an increasing electronic mass renormalization ($Z < 1$). The corresponding effective Hamiltonian to be considered reads:

\begin{equation}
H = -Z \ t \sum_{\langle ij \rangle,\sigma} c^{\dagger}_{i \sigma} c_{j \sigma} - \lambda \sum_{i} n_{i\uparrow} n_{i\downarrow} - \mu \sum_{i} (n_{i\uparrow} + n_{i\downarrow}) \qquad .
\label{resc_H_eq}
\end{equation}
Here the effective bandwidth $D^{*}=2 Z  t$ decreases as we reduce the hole doping and vanishes  at the Mott transition $Z \rightarrow 0$ as $x\rightarrow 0$, while the attractive interaction $\lambda$ is taken as a constant.
Such a simple assumption is explicitly realized, e.g., in realistic modeling of the strongly correlated superconductivity in fullerides\cite{Capone2002,Nomura2015}. As for the energetic balance this amounts to a rescaling in the previous plot, whose effects are reported in the inset of Fig.~\ref{fig:energy_balance}: The results of this a ``more physical'' approach to the problem do not change the qualitative picture, making, however, the energetic balance between the BCS and the BEC regimes overall more symmetric.

\section{DMFT results in presence of a FORCING field}
\label{DMFTwithfield}

% We address the latter study by applying two technically and physically distinct pairing fields. On one side we applied a {\sl uniform} field $\eta$ allowing the pairing term to enter from the beginning of the DMFT loop and self-consistently adjusting the surrounding bath. On the other a {\sl local} field has been adopted. This consists in adding the pairing term to the converged AIM Hamiltonian and performing a single iteration with the fixed equilibrium bath. This procedure inhibits the feedback of the bath and mimics a physical situation in which the single impurity and bath are no longer in thermodynamic equilibrium. The local response may be more relevant to establish a stronger connection with the experimental situation in which the system is locally excited by a laser pulse and the system is driven out-of-equilibrium.   

%After recalling the general physical features of the unperturbed system ($\eta=0$ in Eq. (\ref{complete_H_eq})), 

In this section we will apply a ``theoretical probe'' to investigate the preformed pair physics:  We will study the superconducting response induced  by a {\sl finite} forcing pairing field, also {\sl beyond} the linear response regime.  We will compute by means of DMFT the $s$-wave superconducting order parameter $\Delta = \frac{1}{N} \sum_{i}\langle c_{i\downarrow} c_{i\uparrow} \rangle$ as a function of the external field $\eta$ at different interaction couplings ($U$) and inverse temperatures ($\beta=1/T$).

 %The scope here is to study the superconducting instability effects at different temperature distances from the phase transition $|T-T_C|$ which might strongly affect the superconducting response [\onlinecite{Rohringer2011}].
 
\begin{figure}[h!]
\includegraphics[width=0.45\textwidth]{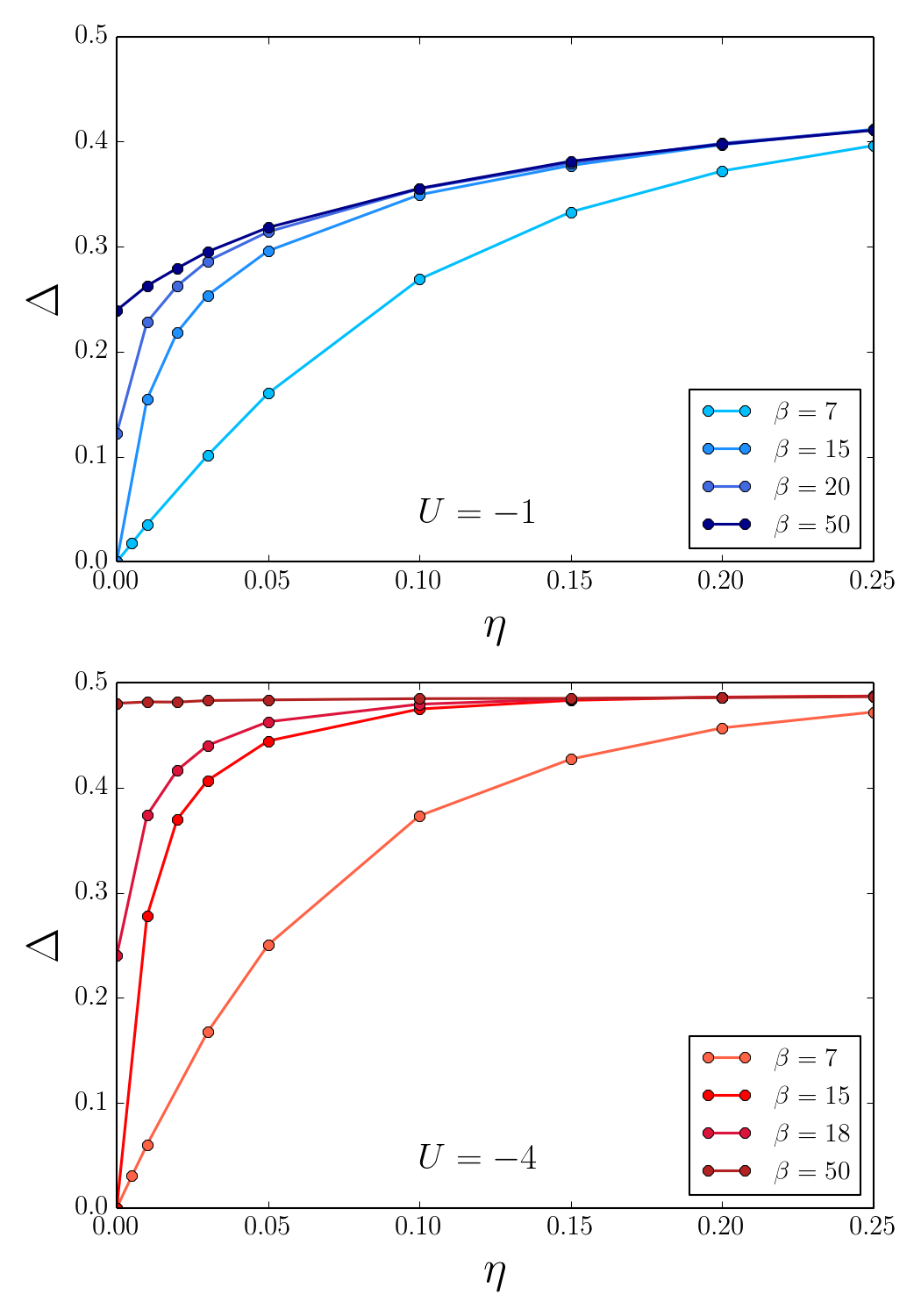}
\caption{Superconducting $s$-wave order parameter for two different values of the on-site attractive interaction, namely: $U=-1$,  and $U=-4$ and for different $\beta$ values. }
\label{fig:gap_diffbeta}
\end{figure}

 As a first step, we follow the evolution of $\Delta(\eta)$ in the attractive case ($U<0$) across the critical temperature  at weak and strong coupling regimes (according to the classification of Sec.~\ref{Model}). This evolution shows the expected appearance of a finite $\Delta$ for $\eta=0$ below $T_c$ and the divergence of the slope of $\Delta(\eta)$ for $\eta \rightarrow 0^{+}$ (which coincides with the linear-response pairing susceptibility) approaching $T_c$ from above (see Fig. \ref{fig:gap_diffbeta}).

%A first inspection of our numerical results shows strong variations of the behavior $\Delta(\eta)$ for small $\eta$, i.e., in the linear response regime. Clearly, these variations mostly reflect the proximity to the onset of a superconducting long-range orded in DMFT, since at $T \rightarrow T_C^{+}$, the slope of $\Delta(\eta)$ becomes infinite at $\eta  \rightarrow 0^{+}$ and the starting value of $\Delta(\eta)=0$ does no longer vanish for $T < T_C$ . These effects of the proximity to the critical temperature $ |T-T_C|$ have been illustrated in Fig. \ref{fig:gap_diffbeta} by choosing four different values of $\beta = 1/k_B T$: one value far below $T_C$ in the superconducting broken phase ($T<T_C$), one right below the superconducting transition ($T\lesssim T_C$), one right above ($T\gtrsim T_C$) and the last one significantly above $T_C$ in the disordered phase ($T>T_C$).  
These obvious features are a direct consequence of a second order phase transition and could, thus, hide the preformed-pair physics. Hence, in the following analysis, we will choose a sufficiently high temperature $T = 1/7 D$ (i.e., $T \gg T_c$ for the selected $U$) to be safely in the normal state and to mitigate the impact of the underlying phase-transition on the low-$\eta$ behavior of $\Delta(\eta)$.  

\begin{figure*}[ht!]
\includegraphics[width=0.45\textwidth]{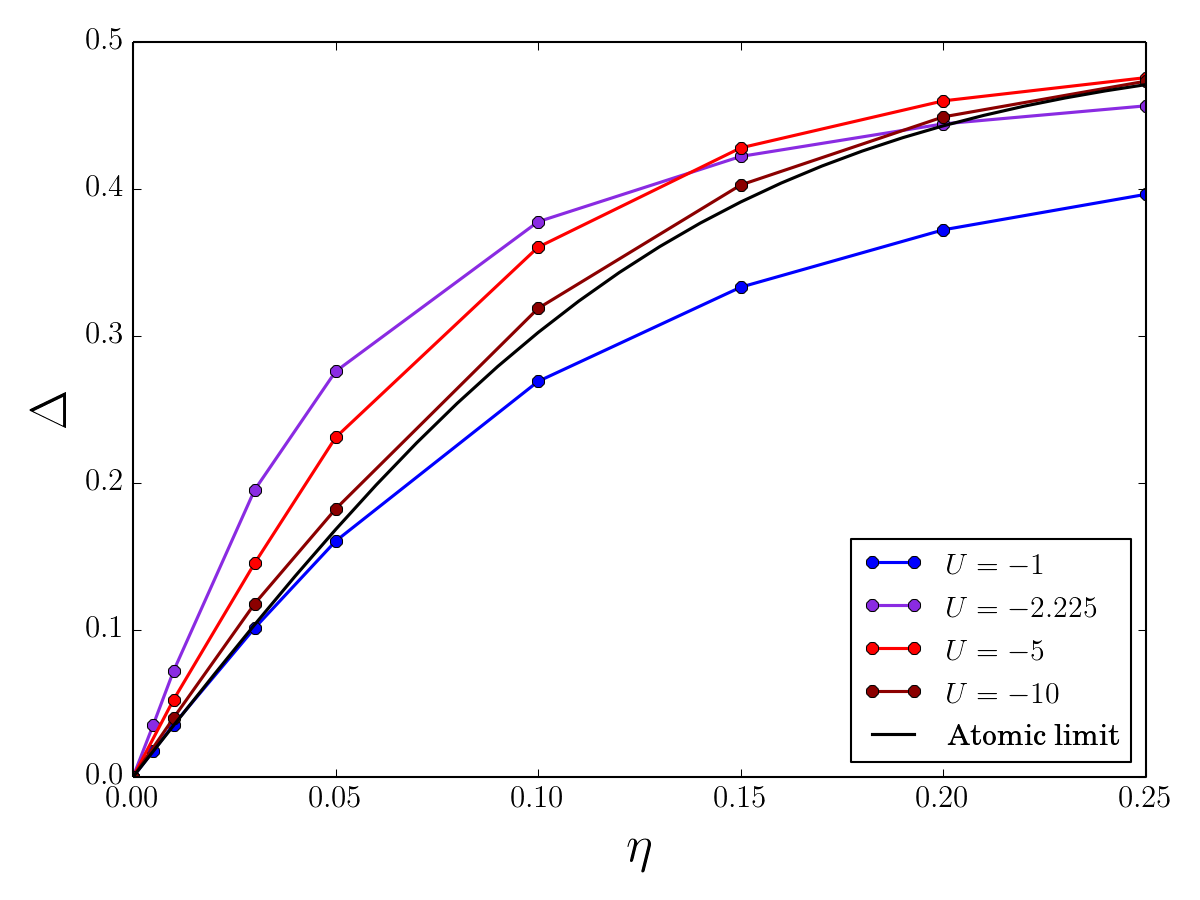} \includegraphics[width=0.45\textwidth, angle=0]{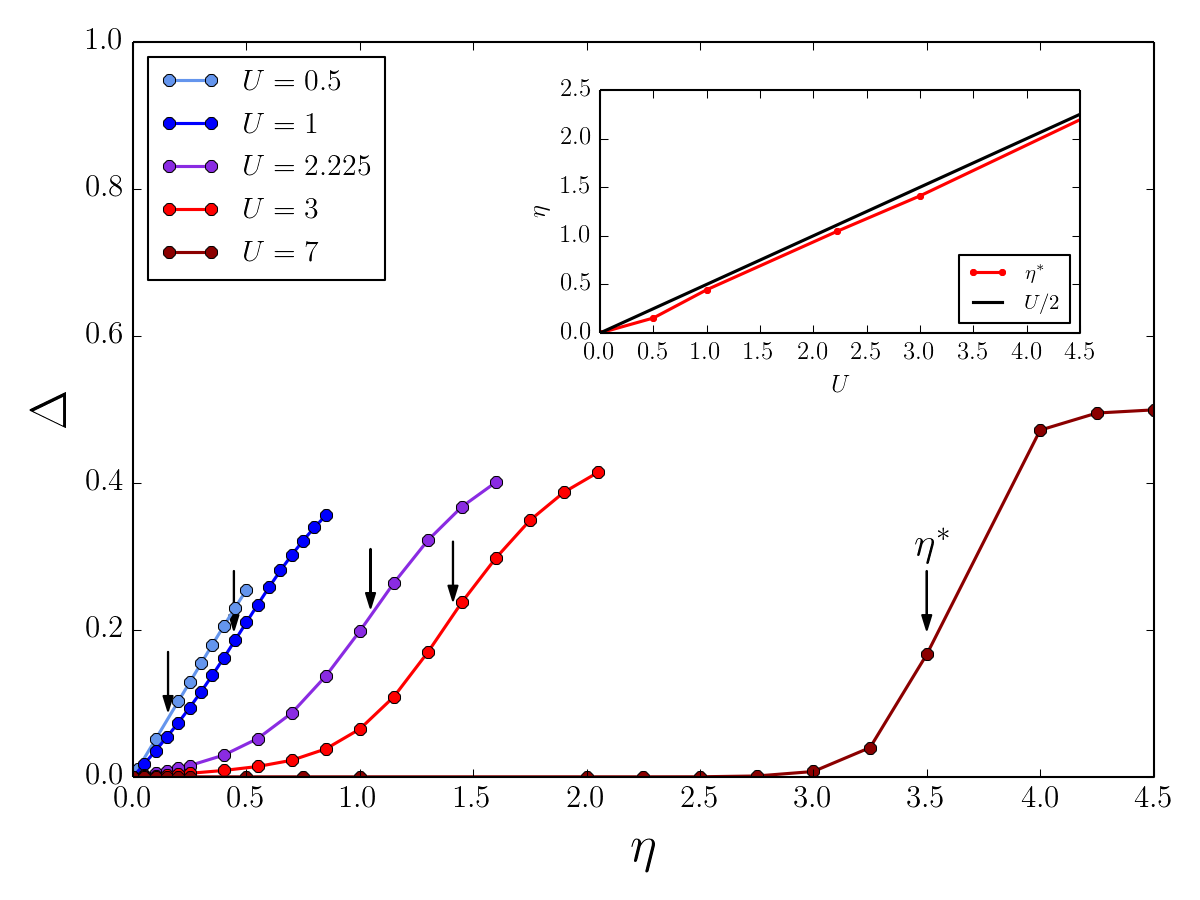} %\qquad
\caption{Superconducting $s$-wave order parameter $\Delta = \frac{1}{N} \sum_{i}\langle c_{i\downarrow} c_{i\uparrow} \rangle$ as a function of the forcing pairing field $\eta$ in the attractive (left panel) and the repulsive (right panel) Hubbard model. Here all the quantities are expressed in unit of the half-bandwidth $D$. The black line on the right panel refers to the analytic behavior in the atomic limit at $\beta=7$ [see Eq.~(\ref{atlim_delta_att})].}
\label{fig:forcing_field}
\end{figure*} 

The corresponding results are shown in Fig.~\ref{fig:forcing_field} (left panel), where the exact result for the atomic limit ($t=0$) is also reported for comparison. For all $U$ values from weak to strong coupling and in the atomic limit $\Delta(\eta)$ saturates to 1/2 by increasing $\eta$. Physically, this reflects the fact that, due to the attractive interaction in the $s$-wave channel, the system responds promptly to the forcing pairing field: the slope of  $\Delta(\eta)$ assumes the largest value for $\eta \rightarrow 0^{+}$ and decreases monotonically with $\eta$.
%, while the saturation value of a complete pairing of the system ($\Delta = 0.5$ for $\eta \rightarrow \infty$ in our units) is gradually achieved. 
Mathematically, this means that $\Delta(\eta)$ is a {\sl concave} function for the whole interval $\eta \in ( 0, \infty)$, i.e.
\begin{equation} 
\frac{d^2 \Delta(\eta)}{d \eta^2} < 0  \; \; \; \forall \; \eta > 0.
\label{second_der_eq}
\end{equation}

\begin{figure}[h!]
\centering
\includegraphics[width=0.40\textwidth]{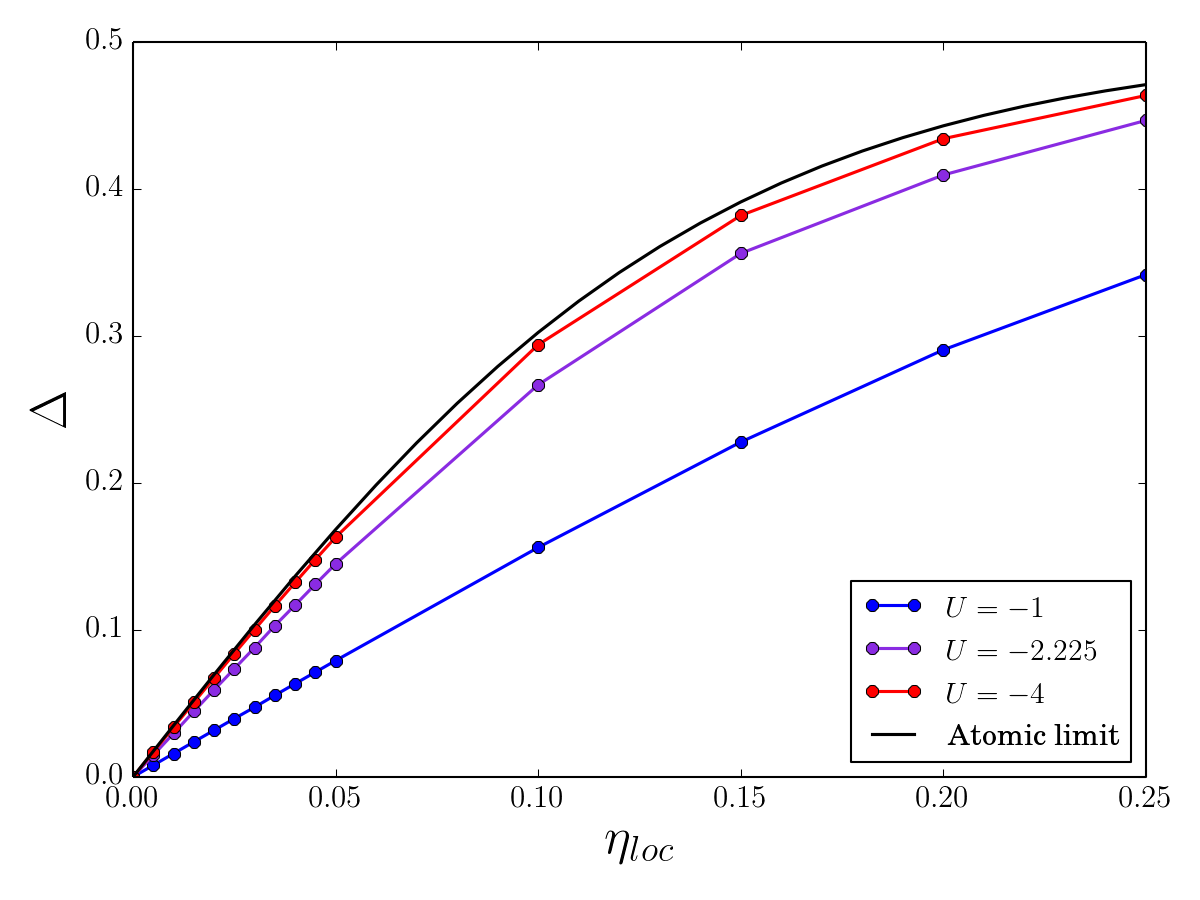}
\caption{Local superconducting $s$-wave order parameter $\Delta = \frac{1}{N} \sum_{i} \langle c_{i\downarrow} c_{i\uparrow} \rangle$ as a function of the local forcing pairing field $\eta_{loc}$ in the attractive Hubbard model. The black line provides the analytic behavior in the Atomic limit case at $U=-7$ and $\beta=7$ [see Eq.~(\ref{atlim_delta_att})].}
\label{fig:forcing_local_field}
\end{figure}

We note that this general property is totally unaffected by the specific behavior of the linear response regime (slope for $\eta \rightarrow 0^{+}$), whose quantitative change as a function of $U$ mostly reflects a different proximity to $T_c$, which is maximum at intermediate coupling\cite{Keller2004,Toschi2005a}, very close to the reported value of $U/D=-2.225$.
This is exemplified in Fig.~\ref{fig:forcing_local_field}  where we perform the same analysis but for a {\sl local} pairing field $\eta_{loc}$ and detecting the local order parameter. Here the small-$\eta$ slope, being  proportional to the local pairing susceptibility, is unaffected by the proximity to the second order phase transition, and it monotonically approaches the atomic limit result. Nevertheless, the curvature of the second order derivative is the same as the one of the uniform-field case.

In order to understand the physical meaning of Eq. (\ref{second_der_eq}), and to exploit it for a preformed-pairs probing beyond our work, we provide a comparative DMFT study of the opposite situation, where our external pairing field $\eta$ contrasts the underlying spontaneously ordered phase of the system at low $T$. 
This can be realized repeating the same analysis for the {\sl repulsive}  ($U > 0$) Hubbard model.  The corresponding results are shown in Fig. \ref{fig:forcing_field} (right panel), where -as before- the {\sl s}-wave superconducting order parameter is plotted as a function of the pairing field $\eta$ for the same high temperature ($\beta=7 D^{-1}$).
The DMFT behavior of $\Delta(\eta)$ shows quantitative and qualitative differences between the attractive and the repulsive case. The first difference concerns the linear response to the pairing field, which is progressively suppressed by increasing the strength of the repulsive interaction.
 
Yet, this difference should be considered -from our perspective- only quantitative: Since the linear response is crucially affected by the proximity to the second order phase transition, it {\sl always} becomes progressively smaller going farther away from the transition (e.g., for the attractive case by increasing $T$, or for the repulsive case by increasing $U$). Hence, its absolute value is not {\sl per se} informative about the presence of preformed pairs in the system.

%\begin{figure}[h!]
%\caption{Superconducting order parameter as a function of the forcing pairing field $\eta$ for different values of $U>0$. The subplot displays the inflection point $\eta^*$ as a function of the interaction strength $U$ (red line) compared with the atomic limit behavior \ref{} (black line) }
%\label{fig:delta_rep_large}
%\end{figure}  

%Much more information, in this respect, can be extracted analyzing the curvature of $\Delta(\eta)$, meaning inspecting its second derivative. 
The behavior of the second derivative of $\Delta(\eta)$, instead, is qualitatively richer than in the attractive case (Fig. \ref{fig:forcing_field}):  For small values of $\eta$, the second derivative has a positive sign (i.e., $\Delta(\eta)$ is a {\sl convex} function), up to an inflection point $\eta^{*}$ (marked by a vertical arrow in the picture). For $\eta > \eta^{*}$ the curvature becomes negative and $\Delta(\eta)$ becomes concave, approaching eventually the regime value $\Delta = 0.5$. %Hence, in the interval $\eta \epsilon \; (0, +\infty)$ the response to an external pairing field qualitative differs in the two systems. 

As a first,  heuristic interpretation of this difference, we observe that the appearance of a region with a {\sl convex} curvature at low $\eta$ reflects the {\sl defiance} of the ($U>0$) system against the formation of the $s$-wave pairs induced by the field. Very large fields instead simply override the repulsive interaction leading to the formation of pairs. 
%Only for large values of the field, the nature of the systems, and specifically, its attitude towards the pairs formation is reverted back to the one of the attractive case, where the external field couples to the pairs already present in the system.
This implies a quite general ``rule-of-thumb": If, by applying a finite pairing field $\eta$ to a system of interest, one observes an initial {\sl convex} curvature of the corresponding superconducting response, the presence of an underlying preformed pair physics (with the same symmetry of the pairing field) can be ruled out. In fact, on the basis of our model results, an inspection of sign changes of the second derivative of $\Delta(\eta)$ should provide a good test for detecting the {\sl absence} of preformed pair physics. This ``rule-of-thumb", which represents one of the main outcomes of our study, will be applied to the more realistic $d$-wave pairing in the pseudogap regime of the Hubbard model in Sec. V.
%Analogous considerations are applicable also to the detection of preformed magnetic moments. Evidently, in this case, the external field to be applied would correspond to a  more usual magnetic field.  
% As for the case of superconductivity, we are mostly interested here, the usage of this rule-of-thumb to the more realistic case of the $d-$wave pairing in the pseudogap regime of the Hubbard model will be discussed at the end of the paper.  

Before proceeding, however, we should recall that only the first derivative of $\Delta(\eta)$ (and, rigorously, only in the limiting case of $\eta \rightarrow 0^+$) has a standard interpretation within the linear response theory.
% and the fluctuation-dissipation theorem. 
Hence, we need to formalize our heuristic understanding of the curvature of $\Delta(\eta)$ more precisely. This will be done in the next section by investigating explicitly, in relevant limiting cases, the relation between the nature of the ground state of the system and its ``defying'' response to the pairing field.

\section{Analysis of limiting cases}
\label{Analyticresults}
Aiming at extracting the physical information encoded in the second derivative of the superconducting order parameter, we perform an investigation of the simplest limiting cases, i.e., non interacting  ($U= 0$), atomic limit ($t = 0$) and the two-site model, where a full analytical treatment is possible.

\begin{figure}[h!]
\centering
\includegraphics[width=0.40\textwidth]{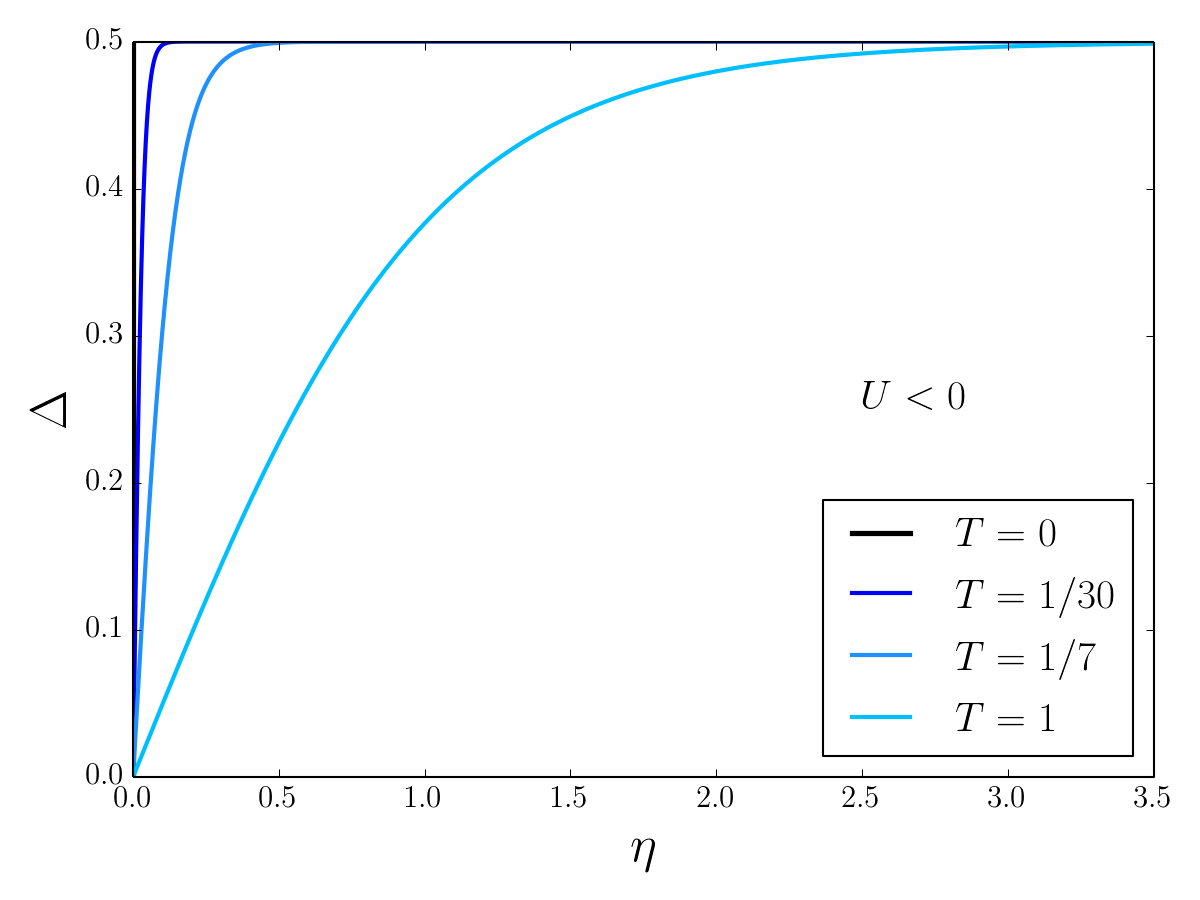}
\includegraphics[width=0.40\textwidth]{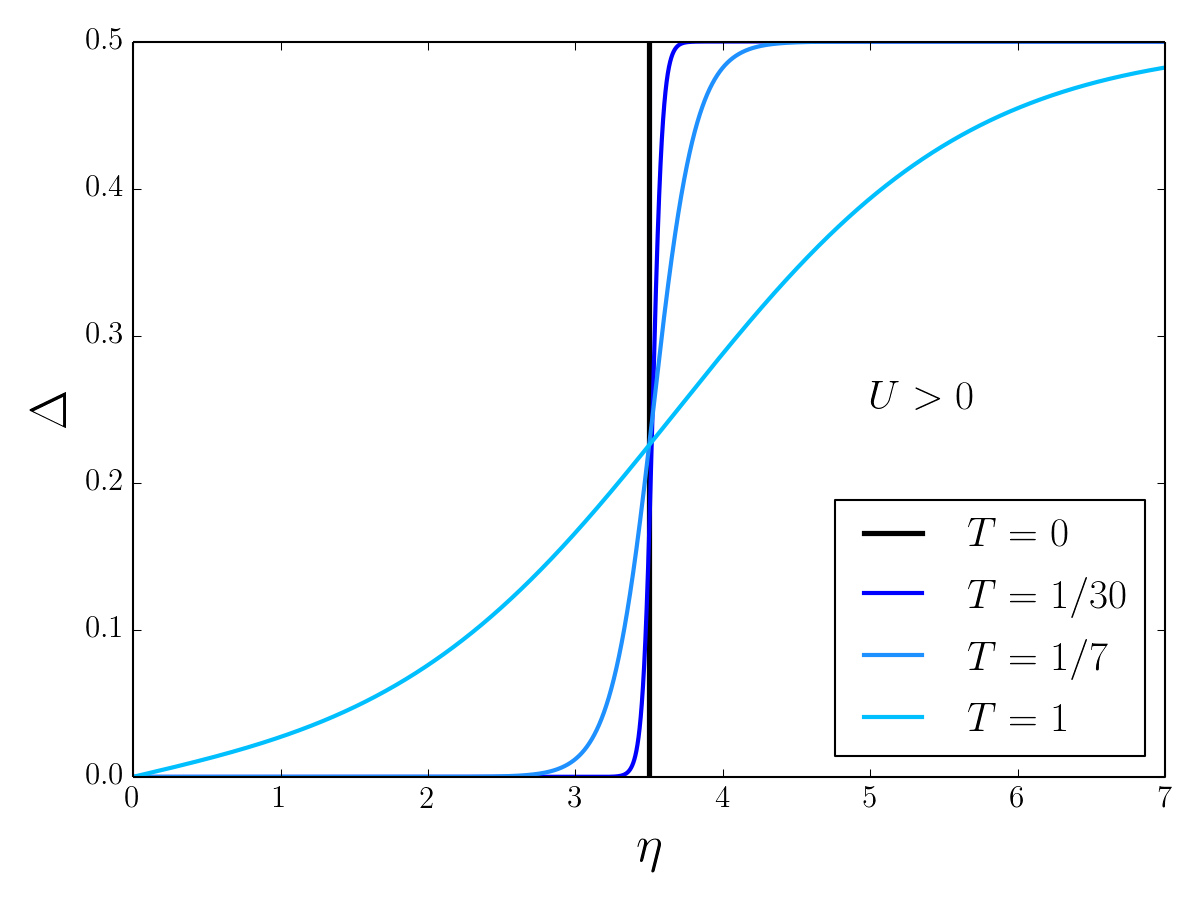}
\caption{Analytic behavior in the atomic limit of the superconducting order parameter $\Delta = \frac{1}{N} \sum_{i}\langle c_{i\downarrow} c_{i\uparrow} \rangle$ with respect to the forcing pairing field $\eta$ in the attractive (upper panel) and in the repulsive (lower panel) Hubbard model for $U=-7$ (and $U=7$ in the repulsive case) and different temperature values (See Eqs. \ref{atlim_delta_att} and \ref{atlim_delta_rep}).}
\label{fig:forcing_field_atomiclimit}
\end{figure}

We start with the non-interacting case which can be diagonalized in momentum space:
\begin{equation}
H = \sum_{\textbf{k} \sigma} \epsilon_{\textbf{k}} c^{\dagger}_{\textbf{k}\sigma} c_{\textbf{k}\sigma} - \eta \sum_{\textbf{k}} (c^{\dagger}_{\textbf{k} \uparrow} c^{\dagger}_{-\textbf{k} \downarrow} + c_{-\textbf{k} \downarrow} c_{\textbf{k} \uparrow})\qquad
\label{noninteracting_H_k}
\end{equation} 
where $\epsilon_{\textbf{k}}$ represents the free-particle energy dispersion.
Because of the presence of the static field $\eta$, one immediately recognizes the formal analogy with the BCS mean-field. After few algebraic steps (see Appendix A), one obtains % the following expression for the {\sl s-wave} superconducting order parameter:
\begin{equation}
\label{noninter_delta}
\Delta (\eta) = \frac{\eta}{\pi D^2}\int_{-D}^{D} d\epsilon \frac{\sqrt{D^2 - \epsilon^2}}{\sqrt{\epsilon^2 + \eta^2}} \  \tanh \Biggl(\frac{\beta \sqrt{\epsilon^2 + \eta^2}}{2} \Biggr) \qquad.
\end{equation}
While this integral can be computed numerically,  it is insighful to study its behavior in the zero and high-temperature regimes for a small pairing field ($\eta \ll 1$ and $\beta \sqrt{\eta^2 + \epsilon^2} \ll 1$). In the first case, we have
\begin{equation}
\label{noninter_lowt}
\begin{split}
\Delta (\eta) \bigg\rvert_{T=0} & = \frac{2\eta}{\pi} \Biggl[\int_{0}^{\eta} d\epsilon \frac{\sqrt{1-\epsilon^2}}{\sqrt{\epsilon^2 + \eta^2}} + \int_{\eta}^{1} d\epsilon \frac{\sqrt{1-\epsilon^2}}{\sqrt{\epsilon^2 + \eta^2}}\Biggr]\\
			  			    & \simeq \frac{2\eta}{\pi}\Biggl[ \frac{\eta^2}{2} + \log(2) -\log(\eta) \Biggr].			  
\end{split}
\end{equation}   
Hence, at $T=U=0$, the first derivative exhibits a positive logarithmic divergence as $\eta \rightarrow 0$, which is readily understood by looking at the non-interacting problem in the limit of vanishing interaction ($U/D \rightarrow 0$). Since $T_c$ decreases exponentially as $U \rightarrow 0$ [\onlinecite{Bardeen1957}], the non-interacting pair susceptibility at $U=0$ must diverge exactly at $T =0$.
In the opposite, high-temperature regime, the expansion of the Fermi function as $\beta \sqrt{\eta^2 + \epsilon^2} \ll 1$ yields:
\begin{equation}
\label{noninter_hight}
\Delta (\eta) \simeq  \frac{\beta}{4} \eta - \frac{\beta^2 \eta}{4\pi} \int_{0}^{1} d\epsilon \sqrt{1-\epsilon^2} \sqrt{\epsilon^2 + \eta^2}\qquad .
\end{equation} 
As the second integral is always positive, we obtain an overall {\sl negative} value of the second derivative of $\Delta$, which vanishes only for $\eta \rightarrow 0$. 
The study of the non-interacting case is not fully conclusive in itself, but it already indicates that a {\sl negative} curvature of $\Delta(\eta)$ does not provide an unambiguous indication of an underlying preformed pair physics, certainly absent in our model for $U=0$. 
%\subsection{Atomic limit}
%\label{atomiclimit}
Further insights can be gained by considering the atomic limit ($t=0$) of Eq.~(\ref{complete_H_eq}). Here, the superconducting order parameter assumes the following expression (see Appendix B):
\begin{equation}
\label{atlim_H}
\Delta(\eta) = \frac{\eta}{2 \ \epsilon} \ \frac{\sinh(\beta \epsilon)}{\cosh(\beta \epsilon) + e^{\beta \frac{U}{2}}} \qquad ,
\end{equation}   
where $\epsilon = \sqrt{\mu^2 + \eta^2} $. In particular, Eq. \ref{atlim_H} further simplifies at half-filling:
\begin{gather}
\label{atlim_delta_att}
\Delta_{\text{att}}(\eta) = \frac{1}{2} \ \frac{\sinh(\beta \eta)}{\cosh(\beta \eta) + e^{-\beta \frac{|U|}{2}}}\\
\label{atlim_delta_rep}
\Delta_{\text{rep}}(\eta) = \frac{1}{2} \ \frac{\sinh(\beta \eta)}{\cosh(\beta \eta) + e^{\beta \frac{|U|}{2}}},
\end{gather}
whose dependence on $\eta$, at different temperatures is plotted in Fig.~\ref{fig:forcing_field_atomiclimit} both for the attractive and the repulsive case.
By exploiting a (rather rough) resemblance of the $\Delta(\eta)$ curves to the corresponding DMFT results of Fig.~ \ref{fig:forcing_field}, we progress in clarifying the physical meaning of the  second derivative of $\Delta(\eta)$. In particular, let us now focus on the repulsive atomic case, whose $\Delta(\eta)$ displays an inflection point at $\eta^{*} = U/2$, in a somewhat similar fashion to the DMFT results. We observe that the inflection point $\eta^{*}$ is exactly associated with a corresponding change of the ground state. By diagonalizing the Hamiltonian (see Appendix B), a crossing of energy levels occurs exactly at $\eta^{*} = U/2$: For $\eta < \eta^{*}$ the lowest energy eigenvalue is achieved in the (degenerate) subspace $\{|\uparrow \rangle, |\downarrow \rangle\}$ describing an (isolated) magnetic moment,  while for $\eta > \eta^{*}$, the ground state becomes   $\frac{|\uparrow \downarrow \rangle + |0\rangle}{\sqrt{2}}$, i.e., a doubly/empty occupied state.
Hence, the curvature of the superconducting response, as a function of the pairing field provides direct information about the ground state properties of the system and, in particular, about the presence or the absence of pairs.
%Specifically,in the atomic limiting case studied here, a positive (negative) curvature of $\Delta(\eta)$ for $\eta \in (0, \eta^{*})$ reflects unambiguously the nature of the ground state of the system, allowing to detect the absence (the presence) of localized electronic pairs induced by the pairing field.

\begin{figure}[h!]
\centering
\subfigure[][]{
\includegraphics[width=0.40\textwidth, angle=0]{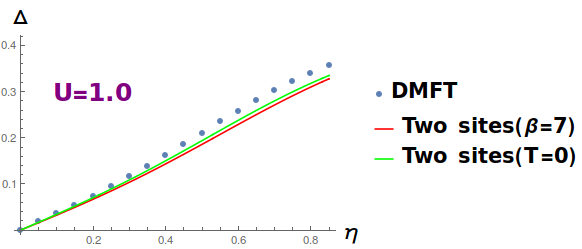}}\\
\subfigure[][]{
\includegraphics[width=0.40\textwidth, angle=0]{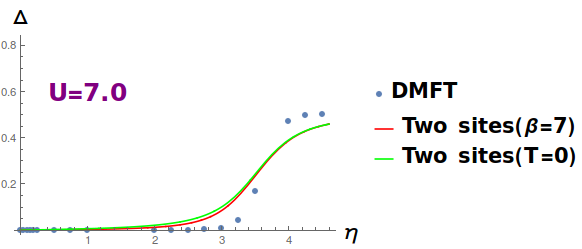}}
\caption{Superconducting order parameter  $\Delta(\eta)= \frac{1}{N} \sum_{i} \langle c_{i\downarrow} c_{i\uparrow} \rangle$ as a fuction of $\eta$: comparison between the DMFT results at $\beta = 7 D^{-1}$ and the two sites results for $T=0$ (Green line) and $\beta = 7 D^{-1}$ (Red line).}
\label{fig:gap_eta_u}
\end{figure}

%\subsection{Two sites model}
%\label{twositesmodel}
The simplest way to verify to what extent these results hold, also for {\sl finite} electron hopping, is to consider a two-site model.
%With this aim, we perform a step beyond the atomic limit, studying the effect of the hopping in a two-site model
%\begin{multline}
%H = -t \sum_{\sigma} (c_{1 \sigma}^{\dagger} c_{2 \sigma} + c_{2 \sigma}^{\dagger}c_{1 \sigma}) + U \sum_{i=1,2} n_{i \uparrow} n_{i \downarrow}\\
 %- \eta \sum_{i=1,2} (c_{i \uparrow}^{\dagger} c_{i \downarrow}^{\dagger} + c_{i \downarrow}c_{i \uparrow})
%\end{multline} 
Here the Hilbert space is spanned by $16$ basis vectors and  the matrix can be readily diagonalized (see Appendix C), allowing to exactly compute the groundstate vector and  $\Delta$ (Fig.~ \ref{fig:gap_eta_u}) as a function of $\eta$.  %As for the former quantity the results for rather different values of $U$ and $T$ are reported in Fig.~ \ref{fig:gap_eta_u}, and benchmarked against the explicit analytical expression for $T=0$ (see Appendix C).  

The agreement with DMFT obviously improves with respect to the atomic limit and an inflection point at $\eta^*$ remains well visible. The broadening around the inflection point is no longer just a mere effect of the temperature as in the atomic case, but it results from the interplay between the magnetic moment and pair formation tendencies of the ground state. Such interplay is -evidently- affected by the hopping term, as it  also happens for the DMFT results.   
%Hence, the analysis of the two-site model provides an improved insight  of the physics encoded in the function $\Delta(\eta)$. 
Also in this case we relate $\Delta(\eta)$, and specifically, the value of its inflection point $\eta^*$, to the corresponding evolution of the system's ground state. For the latter, in presence of the pairing field $\eta$, we obtain
% the following expression:

\begin{multline}
|\text{GS} \rangle = \alpha \Bigl(\frac{|\uparrow, \downarrow \rangle - | \downarrow, \uparrow \rangle}{\sqrt{2}} \Bigr) \\
+ \beta \Bigl(\frac{|\uparrow \downarrow, 0 \rangle + |0,\uparrow \downarrow \rangle}{\sqrt{2}} \Bigr) \\
+ \gamma \Bigl(\frac{|0,0 \rangle + | \uparrow\downarrow, \uparrow \downarrow \rangle}{\sqrt{2}} \Bigr)
\label{twosites_groundstate}
\end{multline}  
where the coefficients $\alpha$, $\beta$, $\gamma$ vary continuously as a function of the field $\eta$ and the interaction strength $U$.  
The first term of Eq.~\ref{twosites_groundstate} represents a singlet-state over the two sites while the two remaining terms feature  empty and doubly occupations.
Intuitively, the singlet state could be linked to the presence of localized magnetic moment (with antiferromagnetic tendency) in the ground state, while the other states contain localized (``preformed'') pairs.  
 
\begin{figure}[h!]
\subfigure[][]
{\includegraphics[width=0.40\textwidth, angle=0]{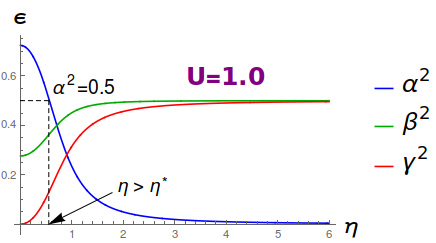}}\quad
\subfigure[][]
{\includegraphics[width=0.40\textwidth, angle=0]{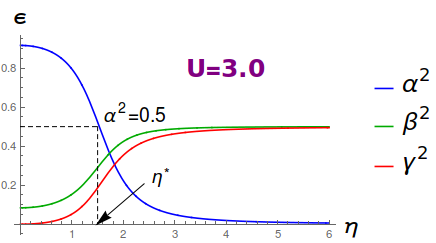}}\\
\subfigure[][]
{\includegraphics[width=0.40\textwidth, angle=0]{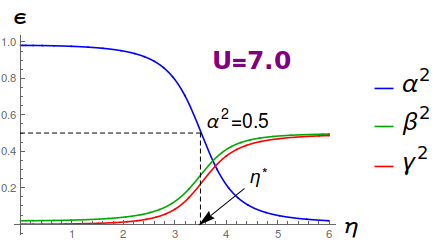}}\\
\caption{Ground state coefficients $|\alpha|^2$ (blue line), $|\beta|^2$ (green line) and $|\gamma|^2$ (red line) of the two-sites model (Eq.~(\ref{twosites_groundstate})) for different $U$ values as a function of the forcing field $\eta$.}
\label{fig:coeff_gs}
\end{figure}

The  explicit dependence of the squared amplitude of $\alpha$, $\beta$, $\gamma$ is shown in Fig.~\ref{fig:coeff_gs} as a function of the pairing field and for different values of the interaction $U$:  The coefficients show a smooth evolution as a function of $\eta$ which becomes sharper  by increasing the  interaction, before recovering the step function of the atomic limit for $U \rightarrow \infty$. For any given value of $U$, $\vert\alpha\vert^2$ monotonically increases with $\eta$, while $\vert\beta\vert^2$ and $\vert\gamma\vert^2$ decrease, reflecting the progressively enhanced weight of ``pairs'' (doubly occupied sites) induced by the pairing field.

By reporting the corresponding values of the inflection point $\eta^{*}$ (as extracted by the data of Fig.~\ref{fig:gap_eta_u})   we find that $\eta^{*}$ occurs in correspondence of the value of $|\alpha|^{2} = 0.5$ (marked by dashed line in the figure) in the intermediate and strong coupling regimes ($U = 3$ and $U = 7$). 
This reflects the fact that, {\sl also} in the two-site model, the sign change of the second derivative of  $\Delta(\eta)$  marks a change of the prevalent character of the ground state. This is dominated by localized magnetic moments ($|\alpha|^2 > 0.5 > |\beta|^2 + |\gamma|^2$) for $\eta < \eta^*$, i.e., where the curvature of $\Delta(\eta)$ is convex. On the other hand, local (``preformed'') pairs prevail ($|\alpha|^2 < 0.5 < |\beta|^2 + |\gamma|^2$) for $\eta > \eta^*$, i.e., where a concave curvature of $\Delta(\eta)$ occurs. Our microscopic analysis confirms thus the link between the curvature of 
$\Delta(\eta)$ 
with the tendency of the system to contrast or to favor the driven superconducting state.

%A clearer analysis of the three coefficients behavior at $\eta^{*}$ can be performed looking at Fig.\ref{fig:coeff_etastar_vsU}.

\begin{figure}[h!]
\centering
\includegraphics[width=0.40\textwidth, angle=0]{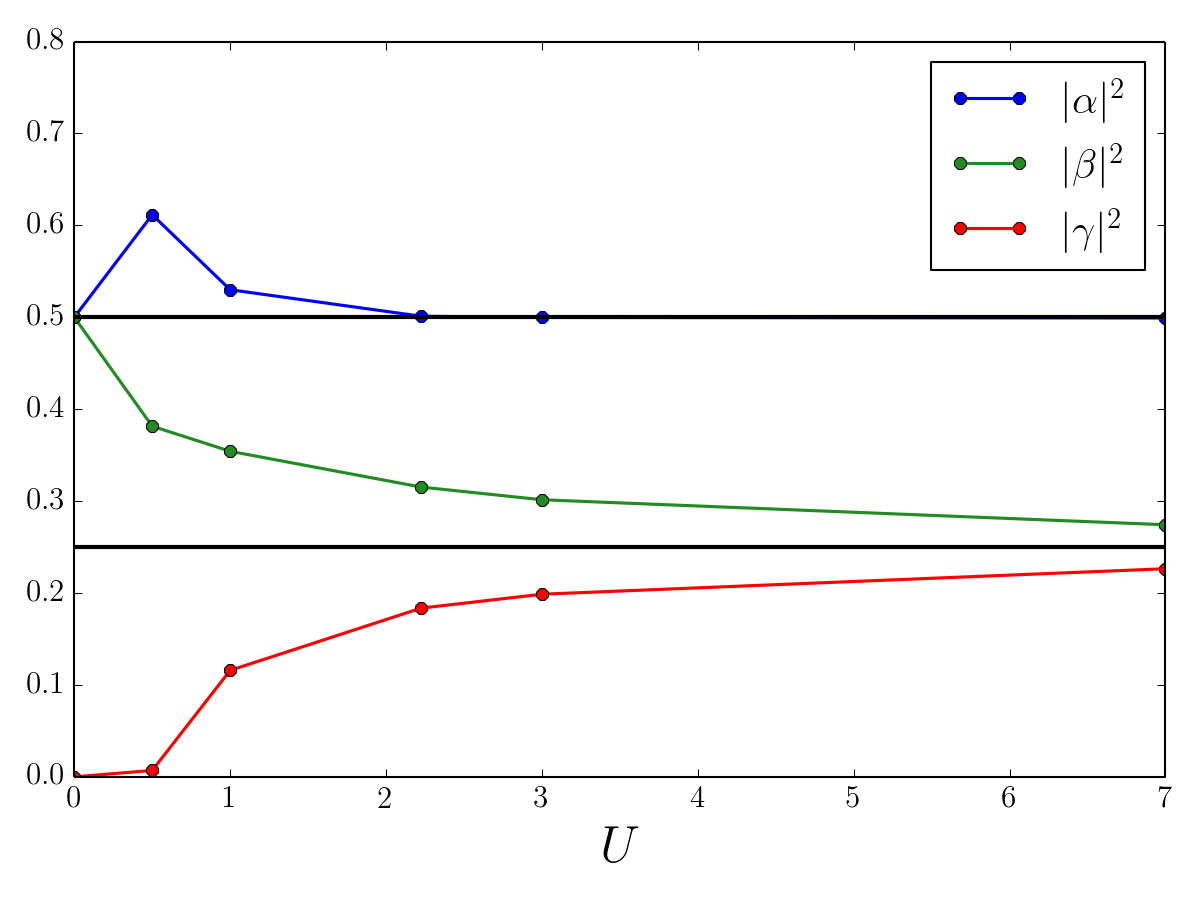}
\caption{ Values of the coefficients $|\alpha|^{2}$, $|\beta|^{2}$ and $|\gamma|^{2}$ of Eq.~(\ref{twosites_groundstate}) at $\eta^{*}$ as a function of $U$.}
\label{fig:coeff_etastar_vsU}
\end{figure}

\begin{figure}[h!]
\includegraphics[width=0.40\textwidth]{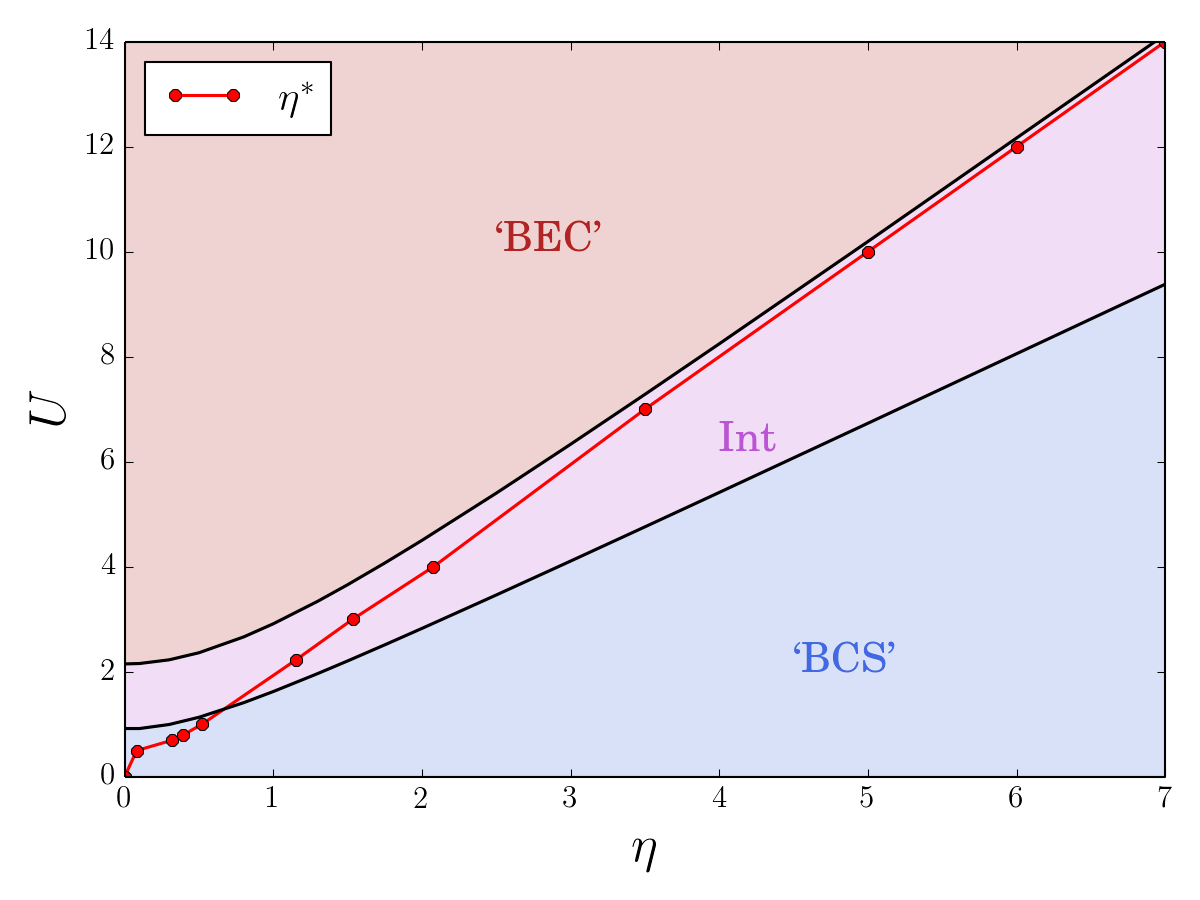}
\caption{ $U-\eta$ plane showing the different regimes obtained from an energy balance analysis in the two-site model. The blue, violet and red regions indicate respectively the weak, intermediate and strong coupling regimes. The $\eta^{*}$ behavior at different $U$ values is shown by the red line.}
\label{fig:etastarvsU_regimes}
\end{figure}

From a quantitative perspective, one should note that in the weak coupling regime ($U=1$, Fig. \ref{fig:coeff_gs}(a)), the inflection point $\eta^*$ is slightly before the coefficient $|\alpha|^2$ crosses the value $0.5$, i.e., at $\eta=\eta^{*}$ one finds an $|\alpha|^2$ slightly larger than $0.5$. While this weak-coupling feature might be a specific result of the two-site model, its presence, however, does not compromise the validity of our interpretation. This can be better understood looking at the Fig.~\ref{fig:etastarvsU_regimes}, where the physics of the two-site model with pairing field is eventually summarized. Here, in a phase diagram $U$ vs. $\eta$ (drawn at a fixed $T=1/7 D$), the values of $\eta^*$, and of the loci where $|\alpha|^2=0.5$ are reported. Moreover, in the spirit of Sec.~II, different region of the phase-diagram could be defined, and classified in terms of the kinetic/potential energy gain/losses induced by the application of the (finite) pairing field:
\begin{eqnarray}
\langle H_{\text{K}} \rangle_{\eta} - \langle H_{\text{K}} \rangle_{\eta = 0}  & = & - 2 \alpha \beta + \frac{2}{U} \Bigl(1+ \frac{4}{U^2} \Bigr)^{1/2}  \\ \nonumber 
\langle H_{\text{pot}} \rangle_{\eta} - \langle H_{\text{pot}} \rangle_{\eta = 0}  & = & - U \alpha^2 - 4 \eta \beta \gamma \\ \nonumber
&  & + \frac{U}{2} \Biggl[ 1 + \Bigl( 1 + \frac{4}{U^2} \Bigr)^{-1/2} \Biggr]  
\end{eqnarray}

The energetic balance analysis of Fig.~\ref{fig:etastarvsU_regimes} confirms, hence, that the correspondence between $\eta^*$ and the change of nature in the ground state of the two-site model is rather solid in the whole intermediate and ``BEC" coupling regime (relevant for the preformed pair physics) with  minor deviations occurring in the ``BCS" regime. 
Moreover, we observe that in the ``BCS" region for $\eta = \eta^*$, $|\alpha|^2$ is slightly larger than $0.5$, indicating that a convex ($=$positive) curvature of $\Delta(\eta)$ is definitely {\sl incompatible} with any preformed pair physics in the ground state (See Fig. \ref{fig:coeff_etastar_vsU}-\ref{fig:etastarvsU_regimes}). We finally note that such criterion of ``absence'', usable to rigorosuly exclude the presence of preformed pairs, is also compatible with the analysis of the concave ($=$negative) curvature of $\Delta(\eta)$ in the non-interacting case, discussed at the beginning of the section.

\section{Implications for other studies}

By analyzing systematically the superconducting response ($\Delta$)  of simple models to an external $s$-wave pairing field ($\eta$), we have demonstrated that, whenever $\Delta(\eta)$ displays a {\sl convex} curvature in the low-field limit, we can safely exclude preformed Cooper pairs. In this section, we discuss the relevance of this ``rule-of-thumb''  criterion to a wider range of systems.
Through a unitary transformation, the pairing field and the superconducting order parameter in the attractive Hubbard model map onto the magnetic field and the magnetic moments in the correspondent repulsive counterpart\cite{BCSBE}. This means that our findings apply also for the case where preformed magnetic moments are driven by an external magnetic field. This generalization of our results is particularly promising, because a measure of the magnetization as a function of the magnetic field is obviously not a pure theoretical probe and it can be directly applied in various experiments (at least for the ferromagnetic case).

%Since  the attractive and the repulsive Hubbard model one onto the other through a unitary transformation which maps the pairing field and the local superconducting order parameter onto a magnetic field and the magnetic moment, our findings apply also for the case where local magnetic moments are driven by an external magnetic field.  This direction is particularly promising because a measure of the magnetization as a function of the magnetic field is obviously not a pure theory probe and it can be directly applied in various experiments (at least for the ferromagnetic case).
%Rather, they are applicable in several experimental set-up: This way the analysis of the induced magnetization ($m(h)$) curvature will also be directly applicable to the experimental results. 
Since the presence and the role of preformed magnetic moments remain a debated issue for several correlated materials, ranging from simple metals like Fe and Ni\cite{Lichtenstein2001,Vonsovskii1993,Anisimov2012,Antipov2012} to alloys (FeAl\cite{Galler2015}) and iron-pnictides and chalgogenides\cite{Yin2011,Toschi2012}, the novel, clear-cut criterion proposed in this work may find widespread application in future experimental and theoretical studies of these materials. 

One may also envisage further applications of our results as an idealized description of pump-probe experiments on superconductors. There, a transient state with an optical response, which is at least compatible with a superconductor, can be created by impulsive excitations inducing coherent phonon deformations, while leaving the temperature of the electrons unchanged\cite{Fausti2012,Cavalleri_cuprates,Novelli,Cavalleri_C60,Giannetti2016}. Comparing experiments against our idealized calculations, one could analyze to which extent the impulsive excitation can be interpreted as an external field driving superconductivity (obviously in our calculations the driven superconductivity is static, as the external field does not depend on time).

\begin{figure}[h!]
\centering
\includegraphics[width=0.48\textwidth]{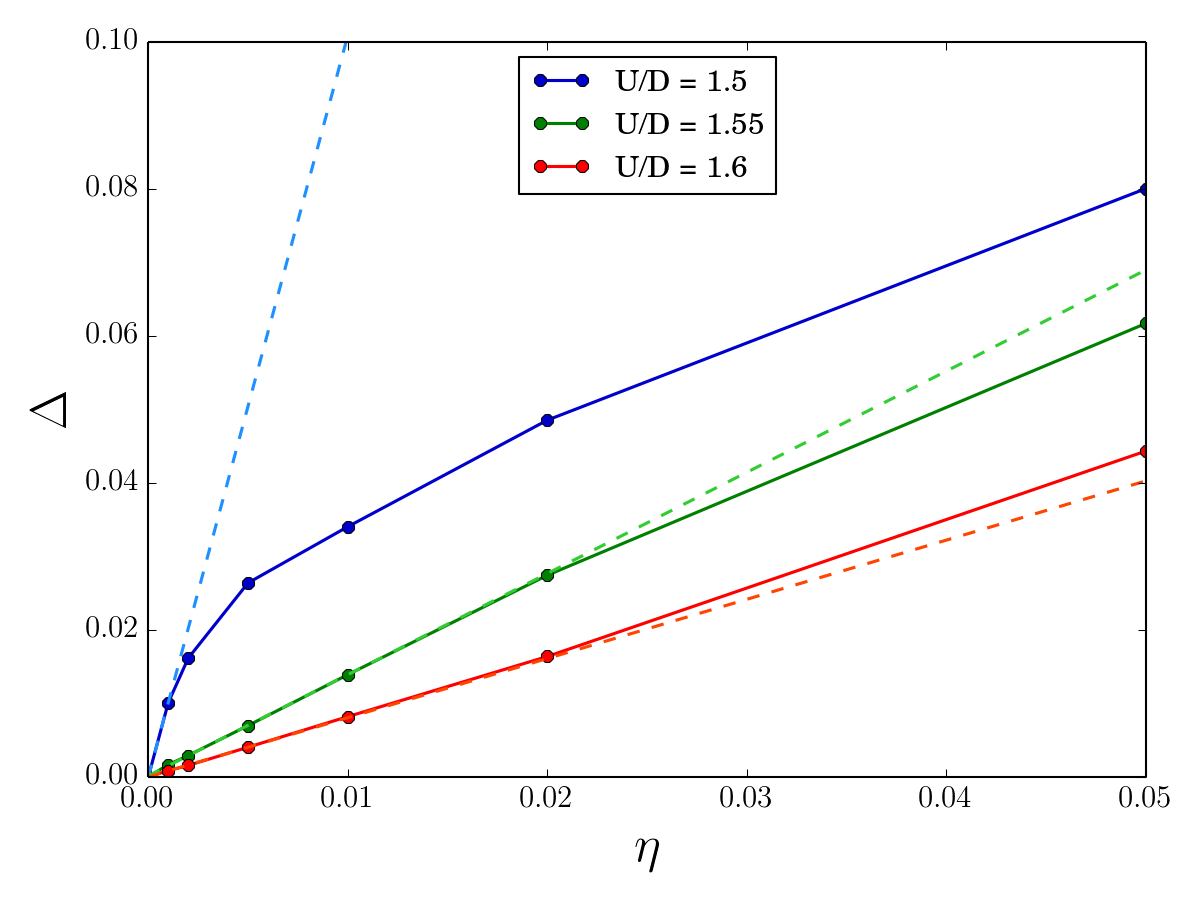}
\caption{[DCA data reproduced from Ref.~[\onlinecite{Gull_Millis2012}]] $d$-wave superconducting order parameter induced by the corresponding $d$-wave pairing field $\eta_{d}$: in a DCA calculation for a 2D Hubbard model at $\beta D = 240$ the order parameter $\Delta_{\textbf{k}}$ is evaluated in the sector $\textbf{k}=(0,\pi)$ and plotted as a function of $\eta_{d}$ at doping $x=0$ for interaction strengths indicated. The dashed lines, marking the slope of the (linear)response at $\eta_d \rightarrow 0^+$, make more easily noticeable the curvature change of $\Delta(\eta_d))$ occurring for the data set at $U=1.6D$.}
\label{fig:gull_millis}
\end{figure}

%Besides applications to the local moment physics, as we anticipated in the Introduction, our results allow to improve the understanding of cutting edge theoretical studies of the pseudogap physics.
One of the most natural applications of our results is, however, the possible presence of preformed d-wave pairs in the pseudogap phase of the two-dimensional Hubbard model. %This, for instance the case of  Dynamical Cluster Approximation (DCA), a cluster extension of DMFT which allows to study non-local effects including d-wave superconductivity and the pseudogap state. 
% studies of the two-dimensional Hubbard model on square lattice (with/without frustration). While the numerical effort of DCA calculations is significantly higher than the DMFT ones presented here, they provide in addition non-local correlations (within a given length scale) as well as the physics of the $d-$wave superconductivity. Hence, within DCA, it is really possible to 
%study the two-dimensional single-band Hubbard model, as the simplest modellization of the cuprates physics, as well as their highly debated pseudogap.
Indeed a closely related theoretical analysis has already been performed using the Dynamical Cluster Approximation (DCA)\cite{DCArev}, an extension of DMFT where the single impurity is replaced by a cluster of $N_c$ sites.  An external $d-$wave pairing field was applied and the resulting $d-$wave superconducting response $\Delta_{\bf k}$ was then computed\cite{Gull_Millis2012}. Without driving fields, for $N_c=8$ and $U > 1.5 D$ (with $D=4t$) the superconducting phase is replaced by the pseudogap state, where a strong spectral weight suppression is found at the antinodal point, without any superconducting long-range order. Clearly,the identification of the physical origin of this phase is also crucial to understand the debated underlying physics of the high-temperature superconductors. In this respect, the two main alternative interpretations describe the pseudogap either as the result of intrinsic interaction effects (spin fluctuations, Mott physics or other) or as the signature of preformed $d-$wave pairs.
% In the perspective of the closely related debate about the origin of the pseudogap in the cuprate experiments, it is crucial to identify the physical origin of the pseudogap state appearing in  the DCA data: This means to define whether, e.g., this stems from (antiferromagnetic) spin fluctuations or it is a non-trivial manifestation of a preformed $d-$wave pseudogap. 
%To this aim, the Authors of Ref.~[\onlinecite{Gull_Millis2012}] performed a pioneering forcing field study: 
The results of  Ref. [\onlinecite{Gull_Millis2012}]  are reproduced in Fig.~\ref{fig:gull_millis}, which shows $\Delta_{\textbf{k}}(\eta_{\textbf{k}})|_{\textbf{k}= (0,\pi)}$ as a function of the external $d-$wave pairing field field $\eta=\eta_d$. On the basis of these calculations the authors concluded that the superconducting response to a $d-$wave forcing field was ``weak enough" to exclude a preformed pair origin of the pseudogap state in DCA. This statement is certainly reasonable, but it still lacks of formal strength as it is not based on some precise criterion.

%As we discussed at length in the previous sections, the first derivative of $\Delta_{\bf k}(\eta_d)$ does not help to identify this criterion, as it is mostly sensitive to the actual ordering temperature but it does not display signatures of the onset of preformed pairs.
%We emphasize that this question is not merely formal, because the actual exclusion of a underlying preformed  pair state depends precisely on its answer. At the same time, it is also not trivial: How can one classify an overall weakness (or strength) of a whole response behavior $\Delta_{\bf k}(\eta_d)$?  Unfortunately, the most obvious criterion of taking the first order derivative in the $\eta \rightarrow 0^+$ limit, does {\sl not} yield the desired information. Rather, according to the linear response theory the low-field slope of $\Delta_{\bf k}(\eta_d)$ is an indication of the proximity to a second order phase transition (here: to a $d$-wave superconducting long-range order). Evidently, the strenght of such slope (which will be always diverging for $T \rightarrow T_C$ and decreasing at higher temperature)  can {\sl not} be invoked as a demostration for the presence/absence of a preformed pair state.
A closer look to the data of Fig.~\ref{fig:gull_millis} shows that, besides an expected progressive suppression of the linear response moving away from the critical regime, one observes a sign change of the second derivative of $\Delta_{\bf k}(\eta_d)$, which starts displaying a convex curvature for small fields at $U = 1.6 D$. Hence, according to our criterion, we can now safely conclude that, once the system is sufficiently far from the superconducting instability, there are no well defined preformed pairs which couple with the external pairing field in a parameter region where the pseudogap is observed in DCA. This excludes a preformed-pair origin of the pseudogap and it  is rather suggestive of a major role played by the strong antiferromagnetic correlations or by non-local Mott physics.
It is worth to mention that the claim of a lack of preformed pairs of Ref.~[\onlinecite{Gull_Millis2012}] made more rigorous by the present analysis,  is not necessarily in contradiction with the claim, based on a different DCA study\cite{Merino2014} of significant $d$-wave pairing fluctuations also far from $T_c$. The latter result indeed refers to  {\sl short-ranged} fluctuations both in space and time, while the (static) forcing-field analysis focuses on long-lived pairs.
% This is quite different information from that provided by the application of a time-constant forcing field $\eta_d$. To the latter, only a long-lived (i.e.: actually preformed) pairing state will react.
Indeed these results are perfectly compatible in view of the more recent study of Ref.~[\onlinecite{Gunnarsson2015}], where it has been shown that well-defined spin-fluctuations, emerging as the predominant pseudogap mechanism  according to our analysis of the data of Ref. [\onlinecite{Gull_Millis2012}], show up as short range/short lived pairing fluctuating modes, if viewed from the perspective of the particle-particle scattering channel.

\section{Conclusions}
This work has been devoted to the definition of an operative criterion to confirm or exclude the existence of preformed pairs in a non-superconducting state. This goal has been achieved by a systematic DMFT study of the response  to an external $s$-wave pairing forcing field $\eta$ in the controlled situation of a single band (attractive and repulsive) Hubbard model. For strong attractive interactions we are indeed certain of the existence of preformed pair region, while the repulsive model does not host any $s$-wave pair.

By comparing the different regimes, we identified a clear-cut ``rule-of-thumb'' criterion for excluding a preformed pair physics: The latter are certainly absent, if the second derivative of $\Delta(\eta)$ is positive in a finite region of $\eta$ from $0$ to a finite value $\eta^*$ (i.e., $\Delta(\eta)$ is a convex function at small fields). This happens because the convexity of $\Delta(\eta)$ reflects the ``reluctance'' of the systems  to respond to the pairing field. In other words, the positive second derivative indicates that the applied forcing field is not intense enough to {\sl revert} the dominant nature of the ground state, which opposes to the order induced by the probing field.  Only if $\eta$ is enough large to exceed any intrinsic property of the model, the curvature changes sign and becomes concave, leading, eventually, to a saturation to the maximum pairing amplitude. This behavior testifies the absence of any intrinsic preformed pairing.

On the other hand, an overall  concave curvature of $\Delta(\eta)$ is a necessary condition for the existence of preformed pairs, but it is not a sufficient condition. Therefore, the latter should be supplemented with further physical information, such as, e.g., the energetic balance underlying superconductivity, as we also discuss in Sec. II of this paper. 

Our results have potential impact for several different aspects. Our ``rule-of-thumb" can be indeed applied, as showed in the previous section, for improving the interpretation of cutting edge DCA calculations\cite{Gull_Millis2012}, relevant for the cuprate physics, and -in particular- to exclude on a rigorous basis that the pseudogap state of DCA is originated by preformed pair fluctuations. Indeed, a related analysis has been proposed in Ref.~[\onlinecite{Gull_Millis2012}], where, however, the physical interpretation of the results was mostly heuristic.

Moreover, it is quite natural to extend the conclusions of our analysis (and, in particular, our criterion) to cases of other, more physical, forcing fields: This would be, e.g., the case of a (finite) magnetic field exploited to detect preformed magnetic moments via the evolution of the magnetization as a function of magnetic field beyond linear response. Evidently, the latter analysis allows also for direct experimental realizations. 

Finally, we notice that the study of regimes of external perturbation field, far away from the linear response, might provide important complementary information to interpret more challenging non-equilibrium phenomena. Comparing our results  with pump-probe experiments where transient superconductivity is realized by coherently exciting phonon modes, we could study whether, and to which extent, these non-equilibrium phenomena can be interpreted in terms of a light-driven pairing field which favors pair formation. 
\section*{Acknowledgments}
We thank E.~Gull, O.~Gunnarsson, C.~Taranto, G.~Sangiovanni, T.~Sch\"{a}fer and G.~Rohringer for insightful discussions. We acknowledge the support from Austria Science Funds (FWF) through the project I-$610$-N$16$, the sub-project I-$1395$-N$16$ as part of the German Research Foundation (DFG) research unit FOR $1346$.

%In fact, in a recent DCA study, a pairing forcing field with $d-$wave symmetry has been already applied as a {\sl theoretical probe} for preformed pairs with unconventional symmetry. However, in this pioneering application, the interpretation of the results was essentially heuristic. Here, instead, by systematically analysing our DMFT results in situations where the absence/presence of preformed pairs is certain, and by comparing them with analytically solvable asymptotic cases, 

\newpage 

\begin{widetext}
\section*{Appendix}
\subsection*{Appendix A: Non interacting case}
\label{app:nonint}

The half-filled non-interacting problem under a isotropic pairing forcing field $\eta$ can be easily solved in \textbf{k}-space. Here, the Hamiltonian assumes the following form:

\begin{equation}
H = \sum_{\textbf{k} \sigma} \epsilon_{\textbf{k}} c^{\dagger}_{\textbf{k}\sigma} c_{\textbf{k}\sigma} - \eta \sum_{\textbf{k}} (c^{\dagger}_{\textbf{k} \uparrow} c^{\dagger}_{-\textbf{k} \downarrow} + c_{-\textbf{k} \downarrow} c_{\textbf{k} \uparrow})\qquad ,
\label{app:noninteracting_H_k}
\end{equation} 

where $\epsilon_{\textbf{k}}$ represents the free particle energy dispersion. 
This Hamiltonian can be exactly diagonalized by exploiting the Bogoliubov transformations [\onlinecite{Bardeen1957}], which read:

\begin{equation}
\begin{cases}
{\gamma_{\textbf{k}, \uparrow}}^{\dagger} = u_{\textbf{k}} {{c}_{\textbf{k},\sigma}}^{\dagger} - v_{\textbf{k}} {c}_{\textbf{-k},\downarrow}\\
\gamma_{\textbf{-k}, \downarrow} = u_{\textbf{k}} {{c}_{\textbf{-k},\downarrow}} + v_{\textbf{k}} {{c}_{\textbf{k},\uparrow}}^{\dagger}\qquad ,
\end{cases}
\end{equation}

where:

\begin{equation}
\begin{cases}
{u_{\textbf{k}}}^2 = \frac{1}{2} \bigl(1+\frac{\epsilon(\textbf{k})}{E(\textbf{k})}\bigr)\\
{v_{\textbf{k}}}^2 = \frac{1}{2} \bigl(1-\frac{\epsilon(\textbf{k})}{E(\textbf{k})}\bigr) \\
\end{cases}
\end{equation}

and $E(\textbf{k}) = \sqrt{{\epsilon_{\textbf{k}}}^2 + {\eta}^2}$.\\ 
In order to evaluate the superconducting order parameter in real space, one has to perform the expectation value of the annihilation ``pair operator" $b_{\textbf{k}} = c_{-\textbf{k} \downarrow} c_{\textbf{k} \uparrow}$ and sum over the Brillouin Zone:

\begin{equation}
\langle b_{\textbf{k}} \rangle = \langle c_{-\textbf{k} \downarrow} c_{\textbf{k} \uparrow} \rangle = u^{*}_{\textbf{k}} v_{\textbf{k}} (1-\langle {\gamma_{\textbf{k} \uparrow}}^{\dagger} {\gamma_{\textbf{k} \uparrow}} \rangle - \langle {\gamma_{\textbf{k} \downarrow}}^{\dagger} {\gamma_{\textbf{k} \downarrow}} \rangle ) = u^{*}_{\textbf{k}} v_{\textbf{k}} (1-2 f(E_{\textbf{k}}) )\qquad ,  
\end{equation}

where $f(E_{\textbf{k}})$ represents the usual Fermi-Dirac thermal distribution for fermion-like excitations with energy $E_{\textbf{k}}$.
Hence, we finally end up with the following expression for the superconducting order parameter:

\begin{equation}
\Delta = \sum_{\textbf{k}} u^{*}_{\textbf{k}} v_{\textbf{k}} (1-2 f(E_{\textbf{k}})) = \sum_{\textbf{k}} \frac{\eta}{2E(\textbf{k})} (1-2 f(E_{\textbf{k}})) = \frac{\eta}{2}\int_{-D}^{D} d\epsilon \frac{D(\epsilon)}{E(\epsilon)} (1-2 f(E(\epsilon))) \qquad .
\end{equation}

Explicitly substituting the DOS of the Bethe-lattice and the Fermi-Dirac thermal distribution, one gets Eq.~(\ref{noninter_delta}).

\subsection*{Appendix B: Atomic limit}
\label{app:atlimit}
In this section we explicitly derive the expression for the superconducting order parameter $\Delta$ in the atomic limit. We proceed in two steps: first we perform the unitary transformation to map the attractive Hubbard model onto the repulsive one, secondly we project the system onto the new principal axes and evaluate the expectation value $\langle c_{i \downarrow} c_{i \uparrow} \rangle$ in the starting (attractive) system. Note that the analogous procedure can be adopted for the repulsive case just by swopping the $U$ sign.\\
Let us start from the attractive Hubbard model Hamiltonian, properly readjusted to emphasize the particle-hole symmetry:\
  
\begin{equation}
H_{\textit{attr}} = U \sum_{i} \Biggl(n_{i\uparrow} -\frac{1}{2} \Biggr) \Biggl(n_{i\downarrow} -\frac{1}{2} \Biggr)\\
 - \Biggl(\mu - \frac{U}{2} \Biggr) \sum_{i} (n_{i\uparrow} + n_{i\downarrow})\\
  - \eta \sum_{i} (c^{\dagger}_{i\uparrow} c^{\dagger}_{i\downarrow} + h.c) \qquad .
\end{equation}

Here $U < 0$ and $\mu$ is the shifted chemical potential (note $\mu = U/2$ at half-filling).
By performing the unitary transformation to map the attractive Hubbard model onto the repulsive one \cite{BCSBE}:

\begin{equation}
\begin{cases}
c_{i\downarrow} \rightarrow (-1)^{n_i} c^{\dagger}_{i\downarrow}  \\
c_{i \uparrow} \rightarrow c_{i\uparrow} 
\end{cases}
\qquad ,
\label{unitary_transform}
\end{equation}

we end up with the corresponding repulsive Hubbard Hamiltonian:
\begin{equation}
H_{\textit{rep}} = |U| \sum_{i} n_{i\uparrow} n_{i\downarrow}  - \frac{|U|}{2} \sum_{i} (n_{i\uparrow} + n_{i\downarrow}) - \Biggl(\mu + \frac{|U|}{2} \Biggr) \sum_{i} (n_{i\uparrow} - n_{i\downarrow}) - \eta \sum_{i} (-1)^{n_i} (c^{\dagger}_{i\uparrow} c_{i\downarrow} + c^{\dagger}_{i\downarrow} c_{i\uparrow} ) \qquad .
\label{eq:rep_hamiltonian}
\end{equation}
 
Notice that Eq. (\ref{eq:rep_hamiltonian}) is nothing but an half-filled repulsive system with two external magnetic fields applied onto the $z$ and the $x$ axes.
Since the first two terms are invariant under axes rotation, it is convenient to perform a unitary transformation of the operators, projecting the system along the principal axes in the ($x$,$z$) plane.

Diagonalizing the last two terms in the $4$-state Hilbert space for the single site and making the proper transformations, we end up with to the following expression:

\begin{equation}
H_{\textit{rep}} = |U| \sum_{i} n_{i\uparrow} n_{i\downarrow} - \frac{|U|}{2} \sum_{i} (n_{i\uparrow} + n_{i\downarrow}) + \epsilon \sum_{i} (n_{i\uparrow} - n_{i\downarrow}) \qquad ,
\end{equation}
where $\epsilon = \sqrt{\Bigl(\mu + \frac{|U|}{2}\Bigr)^2 + \eta^2}$ is the effective magnetic field resulting from the $\mu$ and $\eta$ terms.
 
In order to evaluate the superconducting order parameter $\Delta$, we need to map the pairing operator $c_{i\downarrow} c_{i\uparrow}$ onto the corresponding repulsive system by projecting it along the the principal axes. We obtain:

\begin{equation}
 \langle c_{i\downarrow} c_{i\uparrow} \rangle \rightarrow \frac{1}{4\eta \epsilon^2} \Biggl[-\frac{\Bigl(\epsilon - \frac{|U|}{2} - \mu \Bigr) }{a^2} \langle c^{\dagger}_{i \uparrow} c_{i \uparrow} \rangle + \frac{\Bigl(\epsilon + \frac{|U|}{2} + \mu \Bigr) }{b^2} \langle c^{\dagger}_{i \downarrow} c_{i \downarrow} \rangle \Biggr] \qquad ,
\label{eq:pair_mapping}
\end{equation}
where $a/b = \biggl( \eta^2 + \Bigl(\mu \pm \frac{|U|}{2} \pm \epsilon \Bigr)\biggr)^{-1/2}$.

Evaluating explicitly the expectation values in  Eq. (\ref{eq:pair_mapping}) one obtains:

\begin{equation}
\Delta(\eta) = \frac{\eta}{2 \ \epsilon} \ \frac{\sinh(\beta \epsilon)}{\cosh(\beta \epsilon) + e^{-\beta \frac{|U|}{2}}} \qquad ,
\end{equation}

Which reduces to the simpler expression in the half-filling case:

\begin{equation}
\Delta_{\text{att}}(\eta) =\frac{1}{2} \ \frac{\sinh(\beta \eta)}{\cosh(\beta \eta) + e^{-\beta \frac{|U|}{2}}} \qquad .
\label{eq:response_analy_attrac}
\end{equation}

Since this expression is typically applicable for large values of $|U|$, one can notice that the order parameter exhibits a weak dependence on the interacting constant. This also sets a minimum value for the slope in the limit $|U|\rightarrow +\infty$ at finite temperature:

\begin{equation}
\lim_{|U|\rightarrow +\infty} \left.\frac{\partial \Delta_{\text{att}}}{\partial \eta} \right|_{\eta=0} = \frac{\beta}{2} \qquad .
\end{equation}

A study on the second derivative $\partial^{2} \Delta_{\text{att}}/{\partial \eta}^{2}$ shows that, for all $\eta > 0$, the superconducting order parameter exhibits a negative curvature in the attractive Hubbard model, as shown in Fig.~\ref{fig:forcing_field_atomiclimit} (upper panel). The corresponding behavior for $U>0$ is shown in Fig.~\ref{fig:forcing_field_atomiclimit} (lower panel).

\subsubsection*{Ground state}

As a final atomic limit analysis in the repulsive case, we can look at the ground state evolution as a function of the external pairing field. The local, repulsive Hamiltonian has the following form:

\begin{equation}
H^{\textit{i}}_{\textit{rep}} = U \ n_{i\uparrow} n_{i\downarrow} - \frac{U}{2} \ (n_{i\uparrow} + n_{i\downarrow}) - \eta \  (c_{i\downarrow} c_{i \uparrow} + c^{\dagger}_{i \uparrow} c^{\dagger}_{i \downarrow})
\end{equation}

Where, for sake of simplicity, we reduced at half-filling.

By diagonalizing the $4\times4$ block matrix, we find three different eigenvalues: $-U/2$ (two-fold degenerate), $\pm \eta$. Therefore, as soon as $\eta \rightarrow U/2$, the ground state of the system abruptly changes from the degenerate subspace to the eigenstate associated with $-\eta$, namely:

\begin{equation}
|-\eta \rangle = \frac{|\uparrow \downarrow \rangle + | 0  \rangle}{\sqrt{2}}
\label{eq:new_gs}
\end{equation}  

\subsection*{Appendix C: Two-site model}
\label{app:twosites}

In this section we target the solution of the repulsive Hubbard model taking into account just two sites. The two sites Hamiltonian reads:

\begin{equation}
H = -t \sum_{\sigma} (c_{1 \sigma}^{\dagger} c_{2 \sigma} + c_{2 \sigma}^{\dagger}c_{1 \sigma}) + U \sum_{i=1,2} \bigl(n_{i \uparrow} - \frac{1}{2} \bigr) \bigl(n_{i \downarrow}-\frac{1}{2} \bigr) - \eta \sum_{i=1,2} (c_{i \uparrow}^{\dagger} c_{i \downarrow}^{\dagger} + c_{i \downarrow}c_{i \uparrow})
\end{equation}

Where $t$ is the hopping integral, $U>0$ is the interaction parameter and $\eta$ represents the external pairing field. Since we are working in the grancanonical ensamble, the Hilbert space is spanned by 16 basis vectors, namely:
\begin{gather}
\lbrace |\uparrow,\downarrow \rangle, |\downarrow, \uparrow \rangle, |\uparrow \downarrow,0 \rangle, |0, \uparrow\downarrow\rangle, |\uparrow,\uparrow \rangle, |\downarrow, \downarrow \rangle \rbrace \Rightarrow \text{ssp n}=1 \\
\lbrace |\uparrow,0 \rangle, |\downarrow,0 \rangle, |0,\uparrow \rangle, |0,\downarrow \rangle \rbrace \Rightarrow \text{ssp n}=0.5 \\
\lbrace |\uparrow, \uparrow \downarrow \rangle, |\downarrow, \uparrow \downarrow \rangle, |\uparrow \downarrow, \uparrow \rangle, |\uparrow \downarrow, \downarrow \rangle \rbrace \Rightarrow \text{ssp n}=1.5\\
\lbrace |\uparrow \downarrow, \uparrow \downarrow \rangle \rbrace \Rightarrow \text{ssp n}=2\\
\lbrace |0,0 \rangle \rbrace \Rightarrow \text{ssp n}=0
\end{gather}

where each line indicates a specific subspace (ssp) characterized by an electron density n.
By exploiting the symmetries of the system one can identify for which states the Hamiltonian is diagonal:

\begin{gather}
\lbrace \frac{|\uparrow \downarrow, 0 \rangle - |0, \uparrow \downarrow \rangle}{\sqrt{2}}, \frac{|0, 0 \rangle - |\uparrow \downarrow, \uparrow \downarrow \rangle}{\sqrt{2}} \rbrace \Rightarrow E_1 = 0\\
\lbrace |\uparrow, \uparrow \rangle, |\downarrow, \downarrow \rangle, \frac{|\uparrow, \downarrow \rangle + |\downarrow, \uparrow \rangle}{\sqrt{2}} \rbrace \Rightarrow E_2 = -U
\end{gather}

While projecting the Hamiltonian on the $3$ dimensional subspace
$\lbrace \frac{|0, 0 \rangle + |\uparrow \downarrow, \uparrow \downarrow \rangle}{\sqrt{2}}, \frac{|\uparrow \downarrow, 0 \rangle + |0, \uparrow \downarrow \rangle}{\sqrt{2}}, \frac{|\uparrow, \downarrow \rangle - |\downarrow, \uparrow \rangle}{\sqrt{2}} \rbrace$ gives the following matrix:

\begin{equation}
\text{M}=
\begin{pmatrix}
-U & -2t & 0 \\
-2t & 0 & -2\eta\\
0 & -2\eta & 0
\end{pmatrix}
\end{equation}

One can demonstrate that this matrix has three distinguished real eigenvalues ($E_{3}$, $E_{4}$ and $E_5$) even if an explicit simple expression cannot be found analytically. Nevertheless, we are able to compute numerically eigenvalues and eigenvectors as a function of the interaction $U$ and the pairing field $\eta$.
The remaining states with an odd average number of particles span two equivalent $4$-dimensional subspaces (one for each spin channel) whose Hamiltonian projection reads: 

\begin{equation}
M_{1} =
\begin{pmatrix}
-\frac{U}{2} & -t & -\eta & 0 \\
-t  & -\frac{U}{2} & 0 & -\eta \\
-\eta & 0 & -\frac{U}{2}  &  t \\
0 & -\eta & t &  -\frac{U}{2}
\end{pmatrix}
\end{equation}

So we obtain the two eigenvalues $E_6$ and $E_7$ (each one four times degenerate) and the associated eigenvectors.

\begin{figure}[h!]
\centering
\subfigure[ ][]{\includegraphics[width=0.40\textwidth, angle=0]{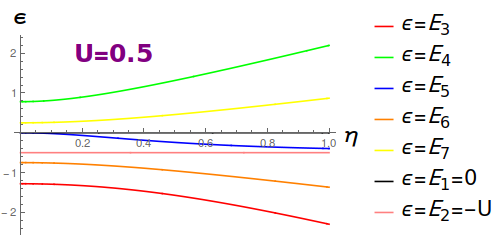}} \quad
\subfigure[][]{\includegraphics[width=0.40\textwidth, angle=0]{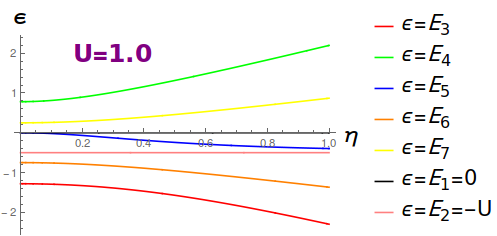}} \\
\subfigure[][]{\includegraphics[width=0.40\textwidth, angle=0]{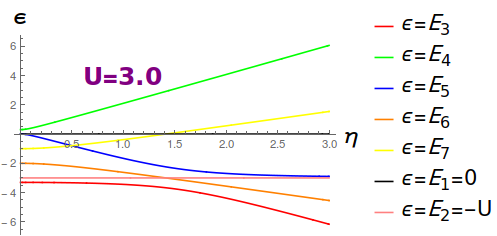}} \quad
\subfigure[][]{\includegraphics[width=0.40\textwidth, angle=0]{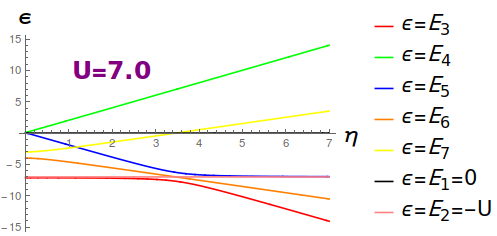}} 
\caption{Energy levels as a function of the pairing forcing field $\eta$ for different values of $U$.}
\label{fig:energy_levels}
\end{figure} 

Fig. \ref{fig:energy_levels} shows the seven eigenvalues as a function of $\eta$ at different interaction values. Hence, it is possible to identify the state associated with $E_{3}$ to be the ground state of the system for any $U$ and $\eta$ values.

This state lives in the subspace described by matrix $M$ and can be written as follows:

\begin{equation}
|\text{GS} \rangle = \alpha \Bigl(\frac{|\uparrow, \downarrow \rangle - | \downarrow, \uparrow \rangle}{\sqrt{2}} \Bigr) + \beta \Bigl(\frac{|\uparrow \downarrow, 0 \rangle + |0,\uparrow \downarrow \rangle}{\sqrt{2}} \Bigr) + \gamma \Bigl(\frac{|0,0 \rangle + | \uparrow\downarrow, \uparrow \downarrow \rangle}{\sqrt{2}} \Bigr)
\label{ground_state}
\end{equation}

\subsubsection*{Superconducting response at $T \neq 0$}

The superconducting order parameter at finite temperature in the two sites model is readily computed as 

\begin{equation}
\begin{split}
\Delta &= =\frac{1}{2} \sum_{i} \frac{1}{Z} \text{Tr}[ e^{-\beta \hat{H}} c_{i\downarrow}c_{i\uparrow}]
		   =\frac{1}{2} \frac{1}{Z} \sum_{i} \sum_{\tilde{n} \ n} e^{-\beta \epsilon_{n}} \langle \tilde{n}|c_{i_\downarrow} |n\rangle (\langle \tilde{n} | c_{i\uparrow}^{\dagger}|n\rangle)^{*}\\
		   &= \frac{1}{Z} \Biggl\lbrace e^{-\beta E_{6}} (m_1 p_1 + m_2 p_2) + e^{-\beta E_{7}} (m_3 p_3 + m_4 p_4) + e^{-\beta E_{3}}(\beta_1 \gamma_1) + e^{-\beta E_4} (\beta_2 \gamma_2) + e^{-\beta E_5} (\beta_3 \gamma_3) \Biggr\rbrace \qquad ,
\end{split}
\label{delta_t}
\end{equation}

where the coefficients $m_{i}$ and $p_{i}$ ( $i=\lbrace1,4\rbrace$) are related to the eigenstates of $M_1$. Namely:

\begin{gather}
|E_{6}\rangle = l_{1(2)} |\uparrow,0\rangle + m_{1(2)} |0,\uparrow \rangle+ n_{1(2)}|\uparrow, \uparrow \downarrow \rangle + p_{1(2)} |\uparrow\downarrow,\uparrow \rangle\\
|E_{7}\rangle = l_{3(4)} |\uparrow,0\rangle + m_{3(4)} |0,\uparrow \rangle+ n_{3(4)}|\uparrow, \uparrow \downarrow \rangle + p_{3(4)} |\uparrow\downarrow,\uparrow \rangle 
\end{gather}

And the partition function is given by:

\begin{equation}
Z = 3 \ e^{\beta U} + 2 + e^{-\beta E_{3}} + e^{-\beta E_{4}} + e^{-\beta E_{5}} + 4 \ e^{-\beta E_{6}} + 4 \ e^{-\beta E_{7}} \qquad .
\end{equation}

Fig.  \ref{fig:gap_eta_u} shows the comparison between equation (\ref{delta_t}) and the DMFT result for the superconducting order parameter.
\end{widetext}
\newpage

%\bibliographystyle{apsrev4-1}
%\bibliography{refs}

%Unused bibitems

%Unused bibitems

\end{document}